\documentclass[3p]{elsarticle}
\usepackage{amsmath,amsfonts,amssymb,graphics,graphicx,color,bm}
\usepackage{graphicx}
\begin{document}
\title{The Lipkin-Meshkov-Glick  model as a particular  limit of the $SU(1,1)$ Richardson-Gaudin integrable models}

\author[slh]{S. Lerma~H.}
\ead{slerma@uv.mx}
\author[jd]{J. Dukelsky}
\ead{dukelsky@iem.cfmac.csic.es}
\address[slh]{Departamento de F\'{i}sica, Universidad Veracruzana, Xalapa, 91000, Veracruz, Mexico}
\address[jd] {Instituto de Estructura de la Materia, C.S.I.C., Serrano 123, E-28006 Madrid, Spain}


\begin{abstract}
The Lipkin-Meshkov-Glick (LMG) model has a Schwinger boson realization in terms of a two-level boson pairing Hamiltonian. Through this realization, it has been shown that the LMG model is a particular case of the $SU(1,1)$  Richardson-Gaudin (RG) integrable models. We exploit the exact solvability of the model to study the behavior of the spectral parameters (pairons) that completely determine the wave function in the different phases, and across the phase transitions. Based on  the relation between the Richardson equations and the Lam\'e differential equations we develop a method to obtain numerically the pairons. The dynamics of pairons in the ground and excited states pairons provides new insights into the first, second and third order phase transitions, as well as into the crossings taking place in the LMG spectrum.
\end{abstract}
\maketitle

\section{Introduction}
\label{intr}
The LMG model was introduced in nuclear physics to mimic the behavior of closed shell nuclei \cite{LMG}. It is a simple model with one quantum degree of freedom. The dimension of  the Hamiltonian matrix increase linearly with the size of the systems allowing its exact diagonalization for large system sizes. As such, the model has been extremely useful to test many-body approximations to nuclear problems (see for example \cite{Schuck} and references therein). More recently the model found applications to many other areas of physics like quantum spin systems \cite{Botet}, ion traps \cite{Unayan},  Bose-Einstein condensates in double wells \cite{Links1} or in cavities \cite {Chen}, and in circuit QED \cite{Larson}. The model has been also utilized to study quantum phase transitions (QPT) \cite{Castanos, duke1} and their relations with quantum entanglement properties \cite{Vidal1, Vidal2}, as well as to explore excited states QPT \cite{Heiss} and quantum decoherence \cite{deco}.
On a different respect, the LMG model was shown to be exactly solvable \cite{Pan} and quantum integrable \cite{NPB707} as a particular limit of the $SU(1,1)$ Richardson-Gaudin integrable models \cite{duke2, RMP}. The most important feature of the exact solution is that it provides a unique form for the wavefunction of the complete set of eigenstates of the model in terms of a set of pair energies or pairons  obtained as a solution of the non-linear coupled Richardson equations. The distribution of pair energies in the energy space change dramatically close to a critical point. A typical example is the exactly solvable $p_x + i p_y$ fermion superfluid derived from the $SU(2)$ hyperbolic RG model \cite{Ger1,Links2,Hyp}. The model has two interesting lines in the phase diagram of density versus coupling constant: a) the Moore-Read line in which all pairons 
 collapses to zero energy, and b) the Read-Green line in which all pairons 
  are real and negative. While in the first case the existence of a QPT is still debated, in the second case it has been shown that Read-Green line corresponds to a third order QPT.

The LMG can be mapped to a two-level boson systems by means of the Schwinger representation of the $SU(2)$ algebra. In this representation the LMG Hamiltonian transforms to a two-level boson pairing model associated with a $SU(1,1)\otimes SU(1,1)$ algebra \cite{Pan}. Within these models, the relation between the distribution of pair energies and the occurrence of a QPT  has been discussed in Ref. \cite{PittDuk} in connection with the {\it s-d} dominance in the Interacting Boson Model of nuclear physics. More recently, a thorough analysis of critical points of the two-site Bose-Hubbard model in terms of the roots of the Richardson equations has been presented \cite{Angela}, showing the intimate relation between quantum criticality and the rapid change in the behavior of the pairons.
 In this paper we will continue these studies focusing on the generalized LMG model which has a rather rich phase diagrams with lines of first and second order QPT and a triple point with a third order QPT. Moreover, the pair energies in a region of the parameter space display a behavior similar to the $p_x + i p_y$ superfluid model, opening the possibility of correlating the physics of both models in the critical regions.

We will start by introducing the LMG model, its Schwinger boson representation leading to a two-level boson pairing model, and the mean field phase diagram of the model in section \ref{LMGbos}. In section \ref{LMG-RG} we will introduce the $SU(1,1)$ RG models and discuss the limits leading to the LMG model. We will introduce in this section a robust numerical method to solve the Richardson equations based on their relation with the Lam\'e ordinary differential equation. Section \ref{GSQPT} is devoted to the study of the behavior of the ground state pairons close to the phase transition, and the region in parameters space which shows a behavior similar to the $p_x + i p_y$ model. Finally, in section \ref{excited} we will describe the RG solutions for the excited states. Concluding remarks are given in section \ref{summa}.

\section{The Lipkin-Meshkov-Glick model and its bosonic representation}
\label{LMGbos}
The LMG model is based on the $SU(2)$ algebra, whose three elements satisfy the commutation relations
$$
[S_{+},S_{-}]=2S_z ~, ~~~ [S_z,S_{}\pm]=\pm S_{\pm}.
$$
The three elements can be considered as the components of the pseudo-spin operator $\bf {S}$. They commute with the Casimir operator of $SU(2)$, ${\bf S}^2=\frac{1}{2}(S_{+}S{-}+S_{-}S{+})+S^2_z$.
In terms of these elements the LMG Hamiltonian can be written as
\begin{equation}
H_{L}=\epsilon S_z+\frac{\lambda}{2}\left(S_+^2+S_-^2 \right)+\frac{\gamma}{2}\left(S_+ S_-+S_-S_+ \right),
\label{LipMo}
\end{equation}
 Note that $H_{L}$ does not commute with the $z$ component $S_z$ for $\lambda\neq 0$ but it  commutes with the total pseudo-spin Casimir operator ${\bf {S}}^2$. Therefore, the Hilbert space of the model can be separated in different sub-spaces labeled by the eigenvalues of the total pseudo-spin  $j(j+1)$, with basis $\mathcal{H}_j=\{ | j m\rangle :  m=-j,-j+1,...,j-1,j\}$. Additionally, the LMG Hamiltonian commutes with the parity operator $\hat{P}=\exp{i\pi (S_z+j)}$, yielding,  for a given $j$, two  invariant sub-spaces ($P=+$ and $P=-$), which are spanned, respectively, by the   basis    $\mathcal{H}_{j+}= \{ | j m\rangle : m=-j,-j+2,-j+4,...\}$  and  $\mathcal{H}_{j-}= \{ | j m\rangle : m=-j+1,-j+3,-j+5,...\}$.   From now on and for  the sake of simplicity, we will assume integer values for $j$.  The semi-integer case can be worked out following the same lines with some slight modifications. For integer $j$ the dimensions of the invariant subspaces, $\mathcal{H}_{j+}$ and $\mathcal{H}_{j-}$, are $j+1$ and $j$ respectively.

Having introduced the LMG Hamiltonian in terms of $SU(2)$  operators, a physical realization of the model requires a representation of the algebra either in terms of a collection of spins or in terms of a fermionic or bosonic system. In its original presentation the $SU(2)$ operators were expressed in terms of a collection of $2j$ fermions distributed in two levels, each having a $2j$ fold degeneracy.  Instead we will make use of the Schwinger boson representation of the $SU(2)$ which allows to a simple connection  with the bosonic RG integrable models.

The Schwinger representation of the $SU(2)$ algebra in terms of two bosons is
\begin{equation}
S_z=\frac{ b^\dagger b-a^\dagger a}{2}=\frac{\hat{n}_b-\hat{n}_a}{2},\ \ \ \ S_+=b^\dagger a,
\ \ \ \ S_-= a^\dagger b,
\label{Schwinger}
\end{equation}
with $a$ and $b$ boson operators, satisfying the usual commutation rules $[a,a^\dagger]=[b,b^\dagger]=1$ and $[a,b]=[a,b^\dagger]=0$. Inserting the boson mapping (\ref{Schwinger}) into the Hamiltonian (\ref{LipMo}) the bosonic version of the LMG Hamiltonian reads:

\begin{equation}
H_{L}=\frac{\gamma+\epsilon}{2} b^\dagger b + \frac{\gamma - \epsilon}{2} a^\dagger a + \frac{\lambda}{2}\left(b^{\dagger 2} a^2+a^{\dagger 2}b^2 \right)+ \gamma \left(  b^\dagger a^\dagger a b \right) . 
\label{LMGb}
\end{equation}

Using the Schwinger representation, a basis for the Hilbert space with total pseudo-spin $j$  can be written in terms of boson creation operators as
$
|j m\rangle= |n_a=j-m, n_b=j+m\rangle,
$
where $|n_a,n_b\rangle=\frac{(a^\dagger)^{n_a} (b^\dagger)^{n_b}}{\sqrt{n_a ! n_b !} } |0\rangle$, with $|0\rangle$ the boson vacuum. Note that for a given $j$ the total number of bosons is constant $N\equiv n_b+n_a=2 j$.  Likewise, the positive and negative parity  basis  in the Schwinger representation are given byin"
$$
\mathcal{H}_{j+}= \{ |n_a=2j, nb=0\rangle,|n_a=2j-2,n_b=2\rangle, |n_a=2j-4,n_b=4\rangle,... \}
$$
$$
\mathcal{H}_{j-}= \{ |n_a=2j-1, n_b=1\rangle,|n_a=2j-3,n_b=3\rangle, |n_a=2j-5,n_b = 5\rangle,... \}.
$$

\begin{figure}[t*]
\centering{
\includegraphics
[width=0.5 \textwidth]{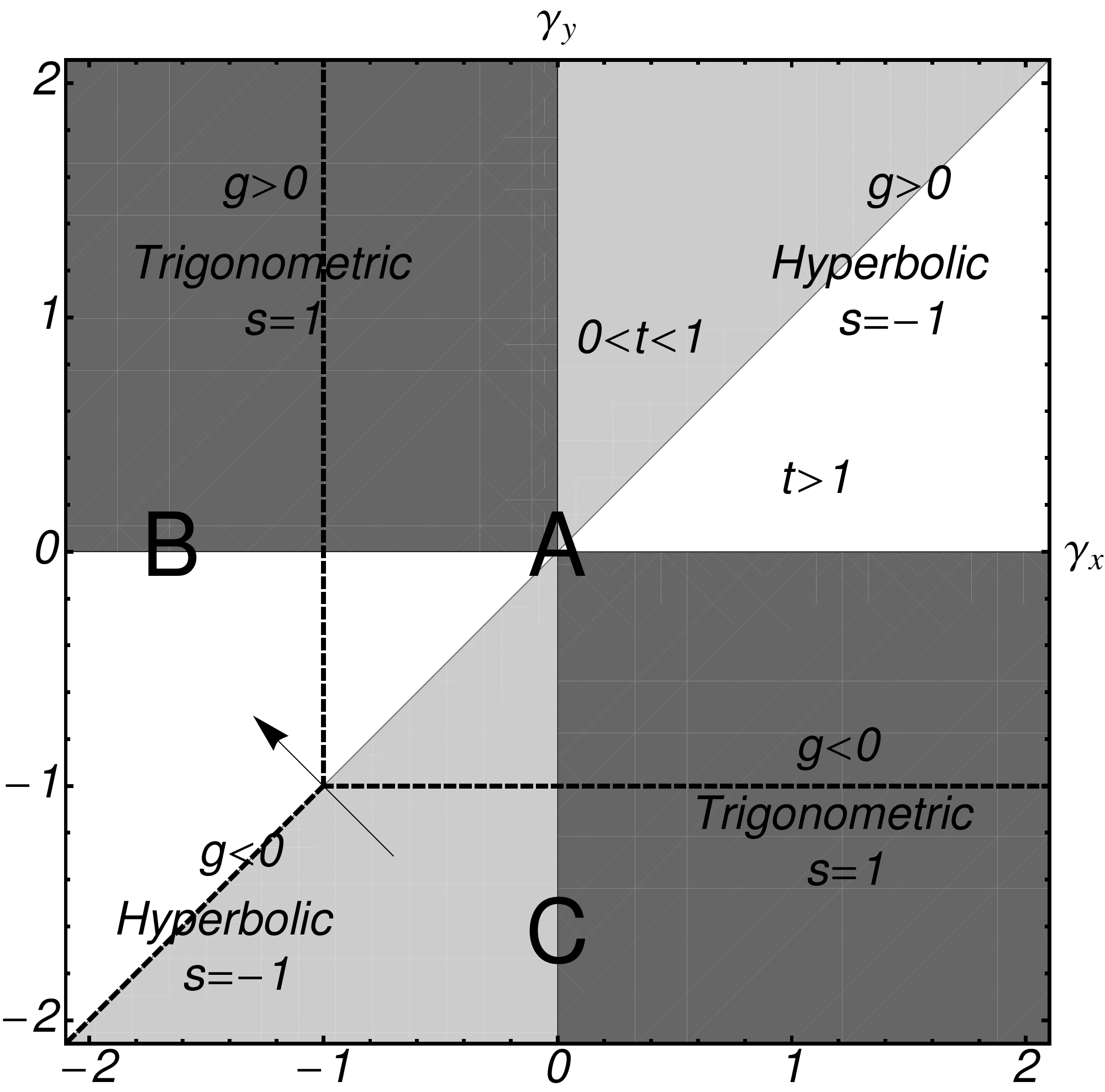}
}
\caption{Phase diagram of the LMG model and Richardson-Gaudin areas of integrability in the $\gamma_x$-$\gamma_y$ parameter space. The  light quadrants correspond to hyperbolic model ($s=-1$) with  light gray for $0<t<1$ and white for $t>1$.  Dark  gray quadrants  correspond to the trigonometric model.  Upper quadrants correspond to positive  $g$ while lower quadrants to negative $g$. The dashed lines separate the three different phases (A,B and C) of the LMG model discussed in the text. The triple point $(-1,-1)$  in the intersection of the lines is a third order transitions when it is traversed in the direction indicated by the arrow.  The horizontal line  $\gamma_y=0$ and the vertical one $\gamma_x=0$  correspond to the rational version of the model ($s=0$). }
\label{fig1}
\end{figure}

 A detailed  analysis of the LMG phase diagram in terms $SU(2)$ coherent states with definite parity has been performed in Ref. \cite{Castanos}. The different phases of the model and the order of their transitions were identified. Here we repeat  that analysis using the Schwinger representation and the boson coherent state

\begin{equation}
|z_a z_b\rangle=e^{-\frac{|z_a|^2+|z_b|^2}{2}} e^{z_a a^\dagger+z_b b^\dagger}|0\rangle,
\label{coher}
\end{equation}
where $z$ are c-numbers parametrized as $z=\rho e^{i \theta}$.
The expectation value of the LMG Hamiltonian  (\ref{LMGb}) in the coherent state $|z_a z_b\rangle=$ is
\begin{equation}
\langle z_b z_a | H_L |z_a z_b\rangle = \frac{\gamma+\epsilon}{2}\rho_b^2 + \frac{\gamma-\epsilon}{2} \rho_a^2 + (\lambda\cos(2(\theta_a-\theta_b))+\gamma)\rho_a^2\rho_b^2.
\end{equation}
The constraint $n_a+n_b=2j$ implies that coherent states parameters should fulfilled $\rho_a^2+\rho_b^2=2j$. Enforcing this relation, the energy surface is
\begin{equation}
E[\rho_b,\theta]=\langle z_b z_a | H_L |z_a z_b\rangle= \frac{2j \epsilon }{2j-1}\left[ A+\Big(2j-1+j B_\theta\Big)\left(\frac{\rho_b^2}{2j}\right)-j B_\theta \left(\frac{\rho_b^2}{2j}\right)^2 \right],
\end{equation}
with $A=\frac{\gamma_x+\gamma_{y}- 2 (2j-1)}{4}$ and $B_\theta=\Big(\gamma_x+\gamma_y+(\gamma_x-\gamma_y)\cos 2\theta\Big)$, where we have used $\theta=\theta_a-\theta_b$, and  the re-scaled parameters defined in \cite{Castanos}
\begin{equation}
(\gamma_x,\gamma_y)\equiv\frac{2j-1}{\epsilon}( \gamma+\lambda,\gamma-\lambda).
\label{gamas}
\end{equation}
 In the thermodynamic limit, $j\rightarrow\infty$, the  energy per particle, $\mathcal{E}[\rho_b,\theta]\equiv E[\rho_b,\theta]/(2j)$,   simplifies to
\begin{equation}
\frac{2\mathcal{E}[\rho_b,\theta]}{\epsilon}+1= \Big(2+B_\theta\Big)\left(\frac{\rho_b^2}{2j}\right) -  B_\theta \left(\frac{\rho_b^2}{2j}\right)^2,
\label{EMF}
\end{equation}
where terms of order $\mathcal{O}(1/j)$ have been neglected.
The phase diagram of the LMG model is obtained by minimizing the energy (\ref{EMF}) with respect to the variables $\theta\in (-\pi,\pi]$ and $\frac{\rho_b^2}{2j}\in[0,1]$ for different values of the model parameters, $\gamma_x$-$\gamma_y$.
The different phases, separated by dashed lines in Fig.\ref{fig1}, are described in Table I where we have classified the  values of the parameters $\theta$ and $\rho_b$ characterizing the coherent state (\ref{coher}) at the absolute minimum. Additionally, Table I shows the energy per particle and the expectation values of the operators $S_x^2$ and $S_y^2$ that play the role of order parameters. The critical line $\gamma_x=\gamma_y<-1$ is a special case because the relative phase $\theta$ is completely undetermined, i.e. the minimum in energy is independent of $\theta$.

\begin{table}
\begin{center}
\begin{tabular}{|c|c|c|c|c|c|c|}
\hline
& & & & &  & \\
 phase& region                 & $\theta_{min}$ & $\left(\frac{\rho_b^2}{2j}\right)_{min}$ & $\left(\frac{2\mathcal{E}}{\epsilon}+1\right)_{min}$ & $\left(\frac{\langle S_x^2\rangle}{j^2}\right)_{min}$ &$\left(\frac{\langle S_y^2\rangle}{j^2}\right)_{min}$\\
& & & & & &  \\
\hline
A&$\gamma_y\geq-1$ and $\gamma_x\geq-1$ &   0&   0& 0 & 0 & 0 \\
B&$\gamma_y>\gamma_x {\hbox{ and }}\gamma_x<-1$ & $0$  or $\pi$  &   $\frac{\gamma_x+1}{2\gamma_x}$ &  $\frac{(\gamma_x+1)^2}{2\gamma_x}$ & $4 \left( \frac{\rho_a^2}{2j}\right)\left( \frac{\rho_b^2}{2j}\right)$ & 0 \\
C&$\gamma_y<\gamma_x {\hbox{ and }}\gamma_y<-1$ &  $\pm\frac{\pi}{2}$ &  $\frac{\gamma_y+1}{2\gamma_y}$  & $\frac{(\gamma_y+1)^2}{2\gamma_y}$. & 0 & $4 \left( \frac{\rho_a^2}{2j}\right)\left( \frac{\rho_b^2}{2j}\right)$ \\
\hline
\end{tabular}
\end{center}
\caption{Phases and their order parameters in the LMG model}
\end{table}

 We are now ready to establish the phase diagram of the LMG model in the Schwinger boson representation. In complete accord with Ref.\cite{Castanos}, three phases are identified, which are distinguished by the occupation of boson $b$ and the relative phase ($\theta$) of the  coherent state parameters $z_a$ and $z_b$.  Alternatively we can characterize the three phases by the order parameters $\frac{\langle S_x^2\rangle}{j^2}= (\rho_a^2/j)(\rho_b^2/j) \cos^2\theta $ and $\frac{\langle S_y^2}{j^2}\rangle=(\rho_a^2/j)(\rho_b^2/j)\sin^2\theta$, where we have neglected terms of order $\mathcal{O}(1/j)$. Phase A ($\gamma_x\geq-1$ and $\gamma_y\geq-1$) has zero occupation of the boson $b$,  $\left(\frac{\rho_b^2}{2j}\right)_{min}= \frac{\langle z_a z_b| \hat{n}_b| z_az_v\rangle}{2j} =0$. Therefore the two order parameters are also zero, $\langle S_x^2\rangle = \langle S_y^2\rangle = 0$. In phase B, ($\gamma_y>\gamma_x$ and $\gamma_x<-1$), the coherent state mixes $a$ and $b$  and the order parameter $\langle S_x^2\rangle/(j^2)$ is finite. Finally, phase C with ($\gamma_y<\gamma_x$ and $\gamma_y<-1$) is the mirror of phase B corresponding to an exchange  between $x$ and $y$. Upon inspection of the order parameters, we can immediately recognize that the transitions between phases (A-B) and (A-C) are continuous in the order parameters, defining a second order phase transition. The transition between B and C is discontinuous in the order parameters characterizing a first order phase transition. These facts were confirmed in Ref. \cite{Castanos} by analyzing the energy derivatives. At the triple point, $\gamma_x=\gamma_y=-1$, both order parameters converge to 0 avoiding the discontinuity of the first order critical line. As shown in \cite{Castanos} this critical point represents a third order phase transition when it is traversed in the direction indicated by the arrow in Fig.\ref{fig1}.

\section{$SU(1,1)$ Richardson-Gaudin  and  LMG models}
\label{LMG-RG}
While the exact solution of the LMG was derived in Ref. \cite{Pan} using an algebraic approach based on the Bethe ansatz, the connection between the LMG Hamiltonian and the  $SU(1,1)$ RG models was  established later \cite{NPB707}.  In order to make the present work self-contained and to fix the notation, we will here derive the exact solution of the LMG Hamiltonian from the more general $SU(1,1)$ RG models following a different path.

The non-compact $SU(1,1)$ algebra, defined in terms of ladder ($K_+, K_-$) and weight ($K_z$) operators  resembles that of the $SU(2)$ group, differing in a sign in the commutation relations
$$
\left[ K^z, K^\pm \right]=\pm K^\pm, \ \ \  \left[ K^+, K^- \right]=-2 K^z .
$$
 Let us now  consider $N_c$ different copies of the $SU(1,1)$ algebra, and construct  $N_c$ linear and quadratic hermitian combinations of the three elements of the algebra
 \begin{equation}
R_i=K_i^z-2 g \sum_{j\not= i} \left[\frac{X(t_i,t_j)}{2} \left( K_i^+ K_j^-+K_i^-K_j^+\right)- Z(t_,i,t_j)K_i^z K_j^z \right],
\label{Ri}
\end{equation}
where $i$, $j$ label each of the $N_c$ copies and $g$ is an arbitrary parameter. The structure of the operators $R_i$ is such that they commute with the total $K^{z}$ operator ($K^z=\sum_{i} K^{z}_{i}$).  It has be shown \cite{duke2}  that the $N_c$ operators commute among themselves ($[R_i,R_j]=0$), defining and integrable model, if the functions $X(t_i,t_j)$ and $Z(t_i,t_j)$ are anti-symmetric functions of an arbitrary set of parameters $t_i$ 
 \begin{equation}
X(t_i,t_j)=\frac{\sqrt{(1+s t_i^2)(1+s t_j^2)}}{t_i-t_j}, \ \ \ \  Z(t_i,t_j)= \frac{1+s t_it_j}{t_i-t_j}.
\label{XZ}
\end{equation}
The parameter $s$ can take three different values  $s=0,1,-1$, defining  the  rational, the trigonometric, and the hyperbolic families of $SU(1,1)$ RG models respectively.

The LMG model is obtained in the limit of two $SU(1,1)$ copies,  which we will label as $a$ and $b$. In the pair boson representation of the $SU(1,1)$ algebra, the elements of the two copies are
\begin{eqnarray}
K_a^+=\frac{1}{2}a^\dagger a^\dagger & K_a^-,=\frac{1}{2}a a, & K_a^z=\frac{1}{2}\left(a^\dagger a+\frac{1}{2}\right)\nonumber \\
K_b^+=\frac{1}{2}b^\dagger b^\dagger & K_b^-,=\frac{1}{2}b b, & K_b^z=\frac{1}{2}\left(b^\dagger b+\frac{1}{2}\right).
\label{su11}
\end{eqnarray}
The irreducible representations (irreps) of the non-compact $SU(1,1)$ algebra are dimensionally  infinite, but they possess a minimum weight state defined by   $K_i^- |MW\rangle=0$. For the previous bosonic representation, these states  are given by $|\nu_i=0\rangle\equiv |0\rangle_i$ and $|\nu_i=1\rangle\equiv |1 \rangle_i$,  where $|0\rangle_i$ is the vacuum of bosons $i=a,b$.  The parameters $\nu_i$ are the so-called seniorities of each of the  $SU(1,1)$ copies. The seniority quantum number, $\nu_i$, counts the number of unpaired bosons $i$ and can take only two values $0$ or $1$. If $\nu_i=0$  the number of bosons $i$  ($n_i$) is even, and  odd if  $\nu_i=1$.

Inserting the boson pair representation of the two copies (\ref{su11}) in the integrals of motion (\ref{Ri}) we construct the two integrals of motion of the LMG model. We can verify the the sum of both integrals gives the conserved quantity $K^z$. Taking the difference between both we obtain:
\begin{eqnarray}
R_b-R_a= K_b^z-K_a^z-2 g X_{ba} \left[ K_b^+ K_a^-+K_a^+K_b^-\right]+ 4 gZ_{ba} K_b^z K_a^z\nonumber\\
 = \frac{1}{2} \left(b^\dagger b-a^\dagger a\right)- g\frac{X_{ba}}{2} \left(b^{\dagger 2} a^2 +a^{\dagger 2}b^2 \right)+ g Z_{ba}\left(  b^\dagger b + \frac{1}{2}\right)\left( a^\dagger a+  \frac{1}{2} \right), \nonumber
\end{eqnarray}
with $Z_{ba}\equiv Z(t_b,t_a)$ and  $X_{ba}\equiv X(t_b,t_a)$. Comparing with the LMG Hamiltonian in the Schwinger representation (\ref{LMGb}), one finds the following relation between the LMG model and the integrals of motion of the $SU(1,1)$ RG models
\begin{equation}
H_L = \epsilon(R_b-R_a)-  \frac{\gamma}{4}, {\hbox{\ \ \ \ \     with  \ \ \ \ \  }} g X_{ba}=-\frac{\lambda}{\epsilon},  \ \ {\hbox{ and }} \ g Z_{ba}=\frac{\gamma}{\epsilon}.
\label{LipRG}
\end{equation}
Without any loss of generality, we choose the parameters entering in $X_{ba}$ and $Z_{ba}$ as $t_b=-t_a\equiv t$, with $t\geq 0$. The functions $X_{ba}$ and $Z_{ba}$ reduce to $X_{ba}=\frac{1+st^2}{2 t}$ and $Z_{ba}=\frac{1-st^2}{2t}$. Then, the relation between the LMG Hamiltonian parameters ($\lambda$, $\gamma$, $\epsilon$) and those of the $SU(1,1)$ RG models are
$$
\frac{\lambda}{\epsilon}=- g \frac{1+st^2}{2t},\ \ \ \ \ \frac{\gamma}{\epsilon}=  g \frac{1-st^2}{2t} .
$$
Or in terms of the $\gamma_x$ and $\gamma_y$ parameters (\ref{gamas}) we have

\begin{equation}
(\gamma_x,\gamma_y)\equiv\frac{2j-1}{\epsilon}( \gamma+\lambda,\gamma-\lambda) = (2j-1)g\left(-st,\frac{1}{t}\right) .
\label{gxgy}
\end{equation}

The relation between the $\gamma_x$ and $\gamma_y$ parameters and those of the RG model classify the quadrants of the phase diagram of Fig.\ref{fig1} in terms of the hyperbolic ($s=-1$) and  trigonometric ($s=1$) RG models. The first ($s=-1$, $g>0$) and third ($s=-1$, $g<0$) quadrants correspond to the hyperbolic RG model, whereas the second ($s=1$, $g>0$) and fourth ($s=1$, $g<0$) are associated  with the trigonometric model. These regions are indicated in  Figure \ref{fig1}, by  dark gray zones  for  the trigonometric model and  light zones for the hyperbolic one. The rational RG model is limited to the $\gamma_x=0$ and $\gamma_y=0$  lines.

The phase transition lines, discussed in the   section \ref{LMGbos} and   shown  in Fig. \ref{fig1} by dashed lines, are translated to the  RG parameters in table \ref{tabla}.

\begin{table}
\begin{center}
\begin{tabular}{|c|c|ccc|}
\hline
transition & line & \multicolumn{3}{c|}{ RG parameters} \\
\hline
first-order&$\gamma_x=\gamma_y$ , $\gamma_x< -1$ & $s=-1$ & $t=1$&  $(2j-1)g= -1$ \\
\hline
second-order& $\gamma_x=-1$,  $\gamma_y>-1$ & $s=+1$& & $(2j-1)g= +\frac{1}{t}$  \\
            &             & $s=-1$ & $t>1$&  $(2j-1)g= -\frac{1}{t} $\\

                          \hline
second-order&$\gamma_x>-1$, $\gamma_y=-1$ &  $s=+1$& & $(2j-1)g= -t$\\
            &                  & $s=-1$& $t<1$&  $(2j-1)g = -t$\\
\hline
\end{tabular}
\end{center}
\caption{Transition lines in phase diagram   $\gamma_x$, $\gamma_y$ and their translation to the RG parameters.}
\label{tabla}
\end{table}

The LMG model has symmetries that relates the spectrum of  systems in two different points in the parameter space. The first of these symmetries  is a point reflection trough  the origin, and relates systems  obtained from a  simple change of sign in the parameters $(\gamma_x, \gamma_y)\rightarrow (\gamma'_x, \gamma'_y)=(-\gamma_x,- \gamma_y)$.  This change in sign is equivalent to a global sign change  in the Hamiltonian, implying that the spectrum of a system is minus the spectrum of the transformed system. In terms of the RG parameter, this transformation corresponds to $g\rightarrow g'=-g$.  A second symmetry of the LMG model is a reflection across the line $\gamma_y=\gamma_x$.  Two mirror points of the phase diagram symmetrically located around this line have exactly the same energy spectrum, as a result of  the invariance of the $SU(2)$ algebra under the  canonical transformation  $S_+\rightarrow -i S_+,~ S_{-}\rightarrow i S_{-},~S_z \rightarrow S_{z}$, corresponding to $b\rightarrow i b, a\rightarrow a$ in the Schwinger boson realization \cite{Van}. In terms of the RG parameters, this symmetry implies that  systems  with parameters related by
\begin{equation}
(g,t) \rightarrow (g',t')=(-sg,1/t),
\label{mirror}
\end{equation}
  have the same spectrum.
  
The $\gamma_y=\gamma_x$ line ($s=-1$ and $t=1$ in terms of RG parameters), located  within the hyperbolic regions of the phase diagram, has the peculiarity that the RG solutions display a singular behavior as it will be shown in   section \ref{GSQPT}. However,  the eigenstates along this line can be easily obtained in closed form  by resorting, for instance,  to the LMG Hamiltonian in terms of pseudospin operators. The condition $\gamma_x=\gamma_y$ implies $\lambda=0$  and  results in the following LMG Hamiltonian
\begin{eqnarray}
H_{L}&=&\epsilon \left(S_z+\frac{\gamma_x}{2(2j-1)}\left(S_+ S_-+S_-S_+ \right)\right)=
\epsilon \left(S_z+g\left(S^2-S_z^2\right)\right),
\end{eqnarray}
which commutes with both $\bf S^2$ and $S_z$, and has eigenstates $|jm\rangle$ and eigenvalues

\begin{equation}
E_m=\epsilon \left(m+\frac{\gamma_x}{(2j-1)}\left(j(j+1)-m^2\right)\right)=\epsilon \left(m+g\left(j(j+1)-m^2\right)\right).
\label{diagE}
\end{equation}
The conservation of the $S_z$ operator along this critical line allows the existence of real crossings between states of the same parity. These crossings take place when $E_{m}=E_{m\pm 2}$. For instance, the $P+$ ground state energy for $g=0$, $E_{m=-j}$, crosses the first $P+$ excited state energy, $E_{m=-j+2}$, when $\gamma_x=\gamma_y=-(2j-1)/(2j-2)$, or  in terms of RG model parameter when $s=-1$, $t=1$ and $g=-1/(2j-2)$.

Having related the two-level $SU(1,1)$ RG models with the LMG model, we can now explore the exact RG solutions in each subspace defined by the number of boson pairs $M$ and seniorities $\nu_a$,$\nu_b$.  For integer $j$ the number of Schwinger bosons, $N=n_a+n_b=2j$, is even,  the seniorities are equal, $\nu_a=\nu_b\equiv\nu=0$ or $1$, and    the  number of boson pairs  is  $M=j-\nu$.
The seniority  sectors $\nu=0$ and $\nu=1$  correspond to the two invariant sub-spaces $P=+$ and $P=-$ respectively.

The eigenvalues $r_i$,  of the $R_i$ integrals of motion are \cite{NPB707}
$$
r_i=d_i\left( 1+ 2 \sum_{\alpha}^M Z(t_i,e_\alpha) + 2\sum_{j\not=i}d_j Z(t_i,t_j)\right),
$$
where $d_i=(1/2)(\nu_i+\frac{1}{2})$, $Z$ is the function defined in (\ref{XZ}), and $e_\alpha$ are the so-called spectral parameters or pairons.  Each particular eigenstate is completely defined by a particular $M$ pairon solution of the coupled set of Richardson equations
$$
1+2\sum_{i} d_i Z(e_\alpha,t_i)+2\sum_{\beta\not=\alpha}^M Z(e_\alpha,e_\beta)=0.
$$

For the particular case of the LMG model, with two $SU(1,1)$ copies, $t_b=-t_a=t$, and  integer $j$  ($\nu_a=\nu_b=\nu=0,1$), the Richardson equations reduce to
\begin{equation}
\frac{1-2gs\left[M+\nu-\frac{1}{2} \right]e_\alpha}{1+s e_\alpha^2}+g\left(\nu+\frac{1}{2}\right)\left(\frac{1}{e_\alpha +t}+\frac{1}{e_\alpha-t}\right)+2g \sum_{\beta\not=\alpha}^M\frac{1}{e_\alpha-e_\beta}=0.
\end{equation}
The eigenvalues of the LMG Hamiltonian are  given by
\begin{eqnarray}
E_L&=&\epsilon(r_b-r_a)-\frac{\gamma}{4}= 
  g\epsilon\frac{(1-s t^2)}{2t}\nu(\nu+1)+ 2g\epsilon \left(\nu+\frac{1}{2}\right)t \sum_{\alpha}\frac{1+s e_\alpha^ 2}{t^2-e_\alpha^2}. \label{enerLip}
\end{eqnarray}
The unnormalized eigenvectors common to the two  integrals of motion ($R_i$) and, consequently, to the LMG Hamiltonian are
\begin{equation}
\prod_{\alpha=1}^M \left(\frac{a^\dagger a^\dagger}{e_\alpha+t}+\frac{b^\dagger b^\dagger}{e_\alpha-t}\right)|\nu_a \nu_b\rangle .
\label{wavfun}
\end{equation}

The Richardson equations can be interpreted as an electrostatic problem in two-dimensions \cite{Elect1, Elect2}. In order to make explicitly this connection we rewrite the Richardson equations as
\begin{equation}
\frac{Q_C}{e_\alpha- P_C}+ \frac{Q_D}{e_\alpha-P_D}+ \frac{\nu+\frac{1}{2}}{2}\left(\frac{1}{e_\alpha+t}+\frac{1}{e_\alpha-t}\right)+\sum_{\beta\not=\alpha}^M\frac{1}{e_\alpha-e_\beta}=0,
\label{RichEqsEl}
\end{equation}
with the effective charges $Q_C$ and $Q_D$
\begin{eqnarray}
Q_C&=&\frac{1}{4g\sqrt{-s}}-\frac{2 j-1}{4}\nonumber \\
Q_D&=&-\frac{1}{4g\sqrt{-s}}-\frac{2 j-1}{4} \label{chargesEff},
\end{eqnarray}
located at position $P_C=-1/\sqrt{-s}$ and $P_D=1/\sqrt{-s}$ respectively.
The pairons have a positive unit charge and they are located at positions $e_\alpha$ in the complex plane. Eq.(\ref{RichEqsEl}) describes the electrostatic interaction of a set  of $M$ pairons with positive unite charge in a two dimensional space. The first two terms in (\ref{RichEqsEl}) describe the electrostatic interaction of the pairons with the two charges $Q_C$ and $Q_D$. For the trigonometric case ($s=1$) the effective charges are  complex $Q_C=-\frac{2 j-1}{4}-\frac{i}{4 g}$, $Q_D=Q_C^*$, and they are located in $P_C=i$, $P_D=-i$. Whereas  in the hyperbolic case ($s=-1$) both, the charges and their positions, are real,  $Q_C=-\frac{2 j-1}{4}+\frac{1}{4g}$ and $Q_D= -\frac{2 j-1}{4}-\frac{1}{4g}$, located in $P_C=-1$ and $P_D=1$ respectively. The third term in (\ref{RichEqsEl}) represents the interaction of the pairons with two charges $\frac{\nu+\frac{1}{2}}{2}$ at positions $\mp t$. Finally, the fourth term corresponds to the mutual repulsion between pairons. Each independent solution of the Richardson equations determines the equilibrium position of the pairons in the complex plane. The electrostatic mapping will be useful to interpret the pairons distribution in each of the quantum phases of the LMG model and the structural changes that take place close to the quantum phase transitions.

\subsection{The Richardson solution as the roots of a generalized Heine-Stieltjes polynomial}

The standard way to solve the Richardson equations is to start from the weak coupling limit where the solution is known [see Eq. (\ref{smallg}) below].   The coupling strength $g$ is increased gradually, using the previous solution as an initial guess to solve the equations for the updated $g$ by means of a standard Newton-Raphson method. A recursive use of this strategy allows to reach the solution for an arbitrary value of $g$ provided one is able to develop a method to treat the numerical instabilities appearing when two or more pairons converge at the same point (at the position of the $t_i$ parameters in this case) generating singularities in the equations \cite{Romb}. Recently, two related methods for solving the Richardson equations have been presented (\cite{solvPan, solvPan2} and \cite{solvGritsev}). Both methods exploit the relation between the Richardson equations and the Lam\'e's Ordinary Differential Equation which has a generalized form of the Heine-Stieltjes polynomials as a solution. The roots of these polynomials are precisely the spectral parameters or pairons. These methods have the advantage of being numerically more stable for system of moderate sizes. For larger systems they have instabilities due to the large precision needed to calculate the roots of a polynomial of high degree. Here we follow the method of references \cite{solvPan, solvPan2}, which is more adequate for systems with a small number of levels \cite{solvLinks2}, as it is the case of the LMG model.

We begin with the Richardson equations in the form (\ref{RichEqsEl}), which can be written as
\begin{equation}
\sum_{\beta\not= \alpha}\frac{1}{e_\alpha-e_\beta}= - \sum_{k=1}^4 \frac{\rho_k}{e_\alpha-\eta_k},
\label{REF}
\end{equation}
with  $(\rho_k, \eta_k)=(Q_C,-1/\sqrt{-s}), (Q_D,1/\sqrt{-s}), ((2\nu+1)/4,-t),$ and $((2\nu+1)/4,t)$ for $k=1,2,3,4$ respectively.

Let us now define the polynomial $P(x)=\prod _{\alpha=1}^M (x-e_\alpha)$ which can be expanded in powers of $x$ as
\begin{equation}
P(x)=\sum_k a_k x^k,
\label{P}
\end{equation}
whose roots are the set  $\{e_\alpha\}$ for a particular solution of the Richardson equations. This polynomial is a generalized Heine-Stieltjes polynomial that satisfies the following Lam\'e's ordinary differential  equation (\ref{app1}):
\begin{equation}
A(x) P''(x)+ B(x) P'(x)-V(x)P(x)=0,
\label{edo}
\end{equation}
where the functions $A,B$ and $V$ are polynomials defined as
\begin{equation}
A(x)= \prod_k^4(x-\eta_k),\ \ \ \ \ B(x)= A(x)\sum_{k=1}^4 \frac{2\rho_k}{x-\eta_k},\ \ \ \ {\hbox{and}}\ \ \ \ V(x)= \sum_{i} 2 \rho_k \Lambda(\eta_k)\prod_{l\not=k}(x-\eta_l),
\label{ABV}
\end{equation}
with
\begin{equation}
\Lambda(x)\equiv \frac{P'(x)}{P(x)}=\sum_\alpha \frac{1}{x-e_\alpha}.
\label{V2}
\end{equation}
The polynomials $A(x)$ and $B(x)$ of degree $4$ and $3$ respectively, depend only on the parameters of the LMG Hamiltonian (\ref{gxgy}). $V(x)$, the so called Van Vleck's polynomial, is at most of third order and  depends on the values $\Lambda(\eta_i)$, which in turn depend on the set of pairons $e_\alpha$,  
\begin{equation}
V(x)=\sum_{i=0}^3 b_i x^i.
\label{V}
\end{equation}
For a general problem  \cite{solvPan}, one can insert the polynomials $V(x)$ (\ref{V}) and $P(x)$ (\ref{P}) in the ordinary differential equation (\ref{edo}) and,  by equating to zero the coefficients at  each order in $x$, one obtains two systems of equations for the coefficients $b_i$ and $a_i$. The first set of equations is linear allowing a solution in which coefficients $b_i$ are expressed in terms of the $a_i$, leaving a second set of non-linear equations for the $a_i$ coefficients. Finally, the $a_i$ coefficients determine the polynomial (\ref{P}) whose roots are the pairons $e_{\alpha}$ of the Richardson equations. For the particular case of the LMG model the first system of linear equations allows to determine the coefficients of the Van Vleck polynomial $b_i$ (except $b_0$)  directly in terms of the parameters of the problem (i.e. they are independent of the $a_i$ parameters). As a consequence, the second set of equations is linear in the coefficients $a_i$.
From Eq. (\ref{ABV}) we obtain:
\begin{eqnarray}
A(x)&=&-s t^2+(s-t^2)x^2+x^4\nonumber\\
B(x)&=& \frac{-s t^2}{g}+\left[ t^2(2j-1)+s(2\nu + 1)\right]x+\frac{s}{g}x^2-2(M-1)x^3.
\end{eqnarray}
After substitution of these polynomials into the differential equation (\ref{edo}), and from the terms of order $M+3$, $M+2$ and $M+1$, we obtain $b_3=0$, $b_2=-M(M-1)$ and $b_1=s M/g$, i.e the Van Vleck polynomial is completely determined except for  the  order zero  coefficient:
$$
V(x)=b_0+\frac{s M}{g}x-M(M-1)x^2.
$$
We can derive the $b_0$ coefficient and the parameters $a_i$ from the orders $0$ to $M$ of the differential equation (\ref{edo}). The result is an eigenvalue equation:
$$
 \sum_{k'=0}^M D_{kk'}a_{k'}= b_0 a_{k} \ \ \ {\hbox {with  }} k=0,1,...,M,
$$
where the matrix $D_{k k'}$ is completely determined by the parameters of the model ($t,g,M,\nu,s$). Its non-zero  matrix elements are given by
\begin{eqnarray}
D_{k\  k-2}&=& (k-2)(k-1-2 M)+M(M-1)\nonumber\\
D_{k\  k-1}&=& s(k-M-1)/g\nonumber\\
D_{k\  k} &=& k((2j-k)t^2+s(2\nu +k))\nonumber\\
D_{k\  k+1}&=& -s(k+1)t^2/g\nonumber\\
D_{k\  k+1}&=& -s(k+2)(k+1)t^2.
\end{eqnarray}
Therefore,  the coefficients $a_k$  of the polynomial $P(x)$  are the elements of each eigenvector of the matrix $D$. Once the coefficients $a_i$ are known, the $e_\alpha$ roots of the Richardson equations are obtained by finding the roots of the polynomial $P(x)$. Each of the $M+1$ eigenvectors of matrix $D$, defines a polynomial whose roots $e_{\alpha}$ correspond to a particular eigenstate of the LMG Hamiltonian.

It is important to remark here that the drawback or bottleneck of the method resides in the last step. As it is well known, finding the roots of high degree polynomials requires a high precision in the determination of the coefficients. Therefore, the number of pairs $M$ is limited to $\cong 10^2-
10^3$. Conversely, the method allows to find directly the pairon roots without resorting to the iterative method of increasing gradually the coupling strength with the burden of having to deal with the singularities of the Richardson equations.

\section{Numerical Results for the ground state}
\label{GSQPT}
In this section we will present and discuss the numerical solution of the Richardson equations for the ground state in the different phases using the trigonometric and the hyperbolic model as required for each particular phase.

\subsection{Trigonometric quadrants}
Ir order to illustrate typical results for the trigonometric regions ($s=1$), we  will consider the line with   $t=1$ as a function of $g$. This line corresponds to $\gamma=0$, which cancels the third term in the LMG Hamiltonian (\ref{LipMo}). The resulting Hamiltonian is the most frequently used in the literature, also known as the Lipkin Hamiltonian. In terms of the scaled parameters this line corresponds to  $\gamma_y=-\gamma_x=(2j-1)g$. For increasing values of $g$, we move along the line from the fourth  to the second quadrant in the phase diagram. According to this diagram, a second order phase transition takes place in the thermodynamic limit  for $\gamma_y=\pm 1$, or equivalently for $g=g_{cr}=\pm 1/(2j-1)$ (see table \ref{tabla}).   The $e_\alpha$ pairons for the ground state as a function of the ratio $g/g_{cr}=g(2j-1)$  are shown in figure 2,  for a system with $j=15$.

\begin{figure}[t*]
\centering{
\includegraphics
[width=0.6\textwidth]{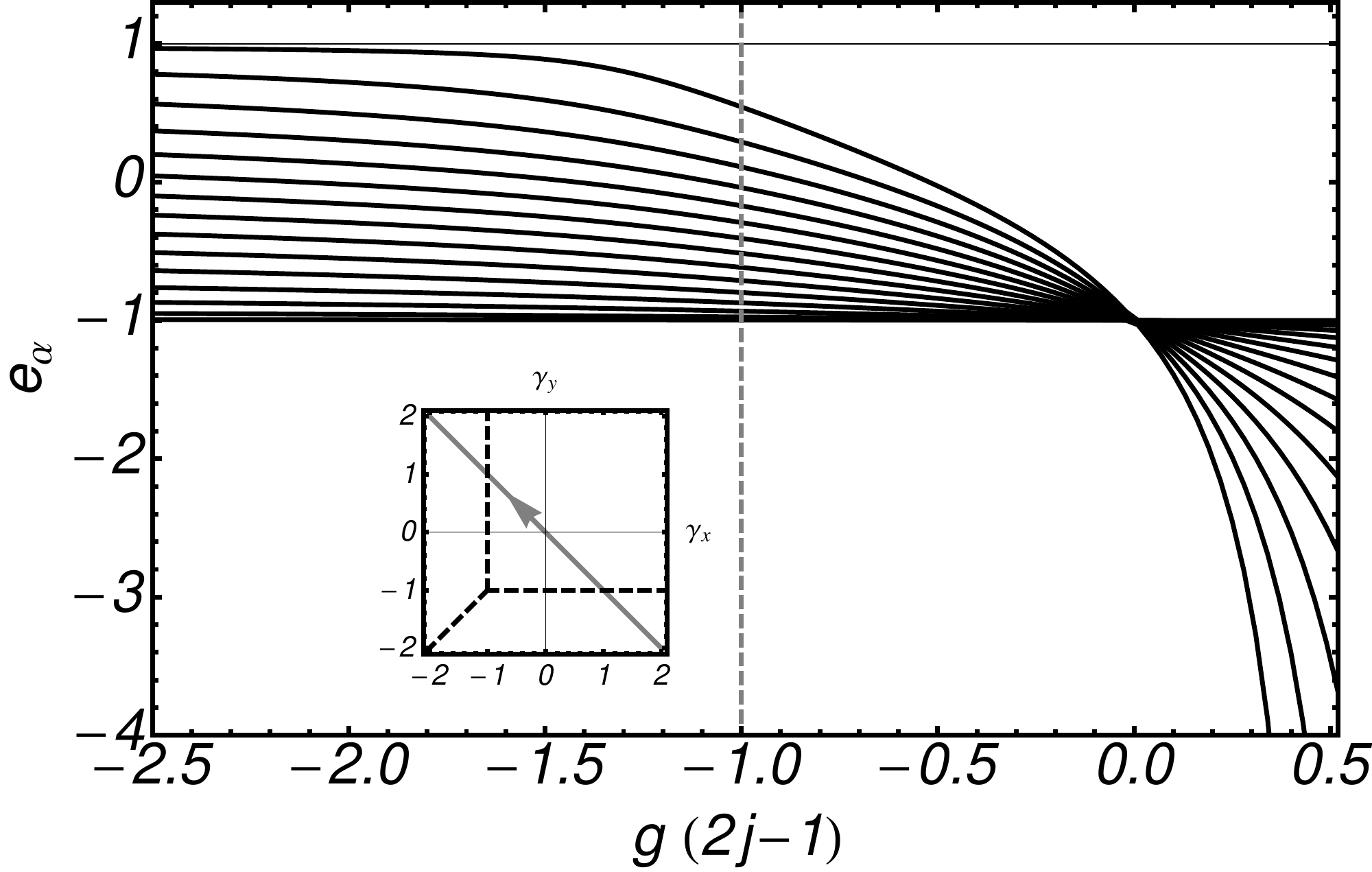}
}
\caption{ Ground state pairons as a function of $g(2j-1)$ for the trigonometric case ($s=1$) with $t=1$ and $j=15$, corresponding to $\gamma_y=-\gamma_x$ as indicated with the arrow line in the inset. Here the arrow corresponds to increasing values of $g$. The vertical dashed line indicates the critical value of $g=g_{cr}=-1/(2j-1)$. Pairons for positive and negative values of $g$ are related by $e_\alpha\rightarrow 1/e_\alpha$.}
\label{fig2}
\end{figure}
As it can be seen in the figure, for  $g\sim 0$ all the pairons are located close to $t=-1$.  As the strength of $g$ is increased they expand in the real axis. For negative $g$, the pairons are constrained to the interval $[-t,t]$, and they behave in a very similar way to the rational case already discussed in the context of the IBM-model \cite{PittDuk}. The second order phase transition can be interpreted as a localization-delocalization transition. The pairons initially localized close to $t=-1$, expand to the entire interval $[-t,t]$ in the transition point.
For  positive $g$  the solutions are, except for a sign in the wave function, entirely equivalent to the negative $g$ case. As it was discussed in  section \ref{LMG-RG},  the Richardson solutions for two mirror points symmetrically located around the $\gamma_x=\gamma_y$ lines [ $(g,t)$ and $(-s g,1/t)$ in terms of RG parameters] have the same spectrum and the wave functions are related by a canonical transformation $b\rightarrow ib$. This symmetry is reflected by a simple relation  between the pairons  with negative and positive $g$   given by $e_\alpha \rightarrow  1/e_\alpha$. It is straightforward to show that  the energy in Eq.(\ref{enerLip}) is invariant under the  transformation $(g,t,e_\alpha)\rightarrow (-s g,1/t,1/e_\alpha)$, and  that this transformation produces a change in the relative sign of the two terms appearing in the product wave function (\ref{wavfun}), in agreement with the canonical transformation  $b\rightarrow ib$.

 In the trigonometric quadrants, the dynamics of the pairons as a function of the control parameters $[\gamma_x,~\gamma_y]$ take place entirely in the real axis and it is very much like the already known dynamics of pairons in the rational boson pairing models \cite{PittDuk}.

\subsection{Hyperbolic quadrants}

The hyperbolic regions ($s=-1$) of the phase diagram offer  much richer structures than the trigonometric regions. In order to illustrate this issue, we will study a system with $j=15$ and $t=1/2$, which corresponds to the line $\gamma_y= 4 \gamma_x$ in phase diagram  of Fig.\ref{fig1}. The line traverses the third and first quadrants from below for increasing values of $g$, and has a critical point of a second order phase transition for $\gamma_{y}=-1$, corresponding to $g(2j-1)=-t=-1/2$ (see table \ref{tabla}).

\begin{figure}[t*]
\centering{
\includegraphics
[width=0.6\textwidth]{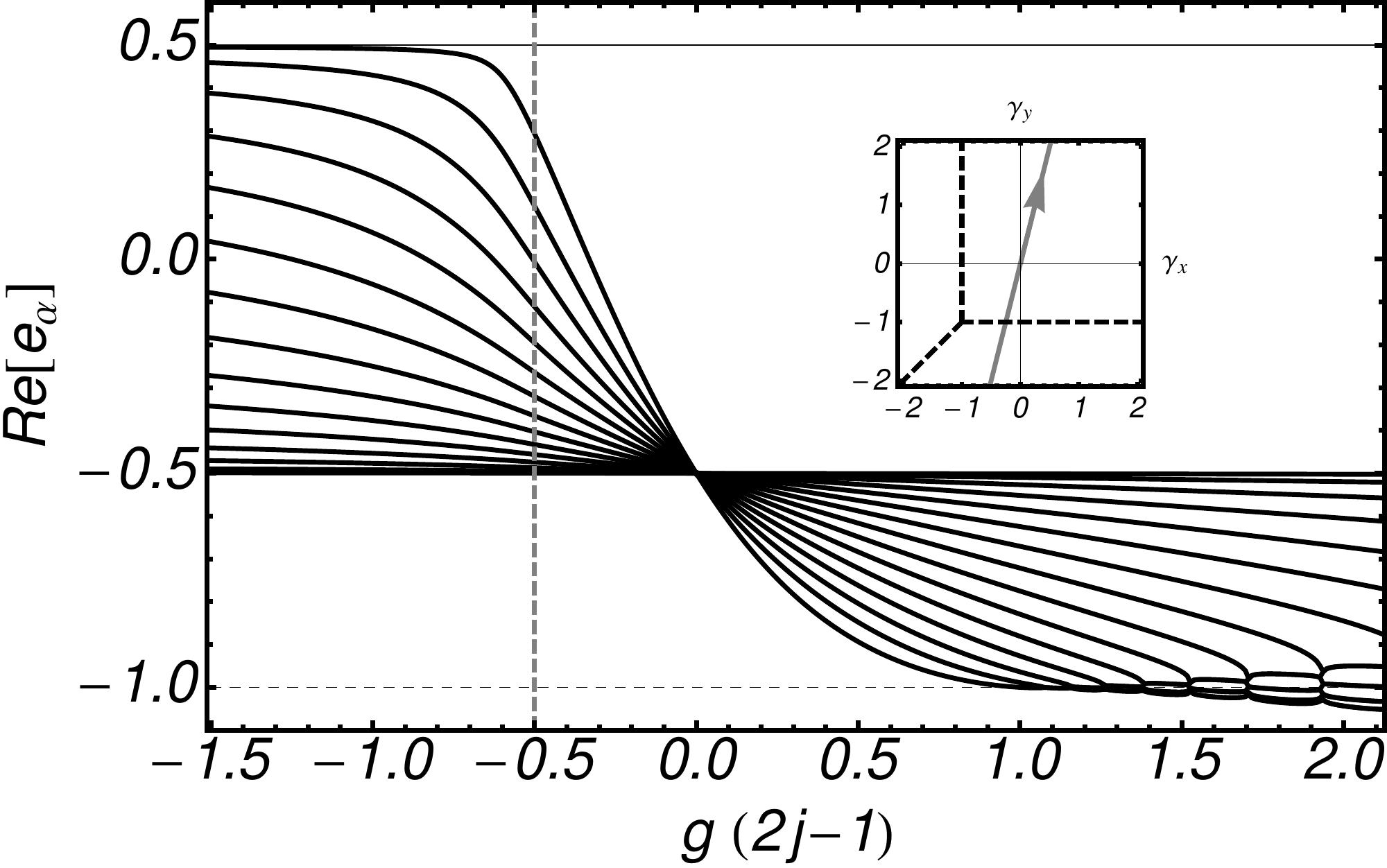}
}
\caption{
Real part of the ground state pairons as a function of $g(2j-1)$ for the  hyperbolic case ($s=-1$) with $t=1/2$ and $j=15$, corresponding to $\gamma_y=4\gamma_x$ as indicated with the arrow line in the inset. The arrow corresponds to increasing values of $g$. The vertical dashed line indicates the critical value of $g=g_{cr}=-1/(4j-2)$ corresponding to the line $\gamma_y=-1$. For positive $g$, successive collapses of pairons in the position $P_C=-1$  (horizontal dashed line) of the effective charge $Q_c$ can be seen.}
\label{fig3}
\end{figure}

Fig. \ref{fig3}   shows the ground state pairon roots as  function of $g(2j-1)$. Similarly to the trigonometric results, the pairons converge to $-t$ for $g\rightarrow 0$. For negative values of $g$ the pairons are constrained  to the interval $[t,-t]$ with the phase transitions (vertical dahsed line)  signaled by the delocalization of the pairons in this interval. For the $g$-positive case (where no phase transtition is expected) an interesting behavior of the pairons takes place. As the coupling $g$ is increased the pairons collapse successively to the position ($P_C=-1$) of the effective charge $Q_C$.  In \ref{app2} it is shown that a necessary condition  to have  $N_C$ pairons collapsing to $P_C=-1$  is
\begin{equation}
g=g^c_{N_C}\equiv\frac{1}{2j+1-2 N_C},
\label{cond1}
\end{equation}
with $N_C=1,...,M$ and   $0<g^c_{N_C=1}<g^c_{N_C=2}<...<g^c_{N_C=M}$.

\begin{figure}[t*]
\centering{
\begin{tabular}{lr}
\includegraphics[width=0.415\textwidth]{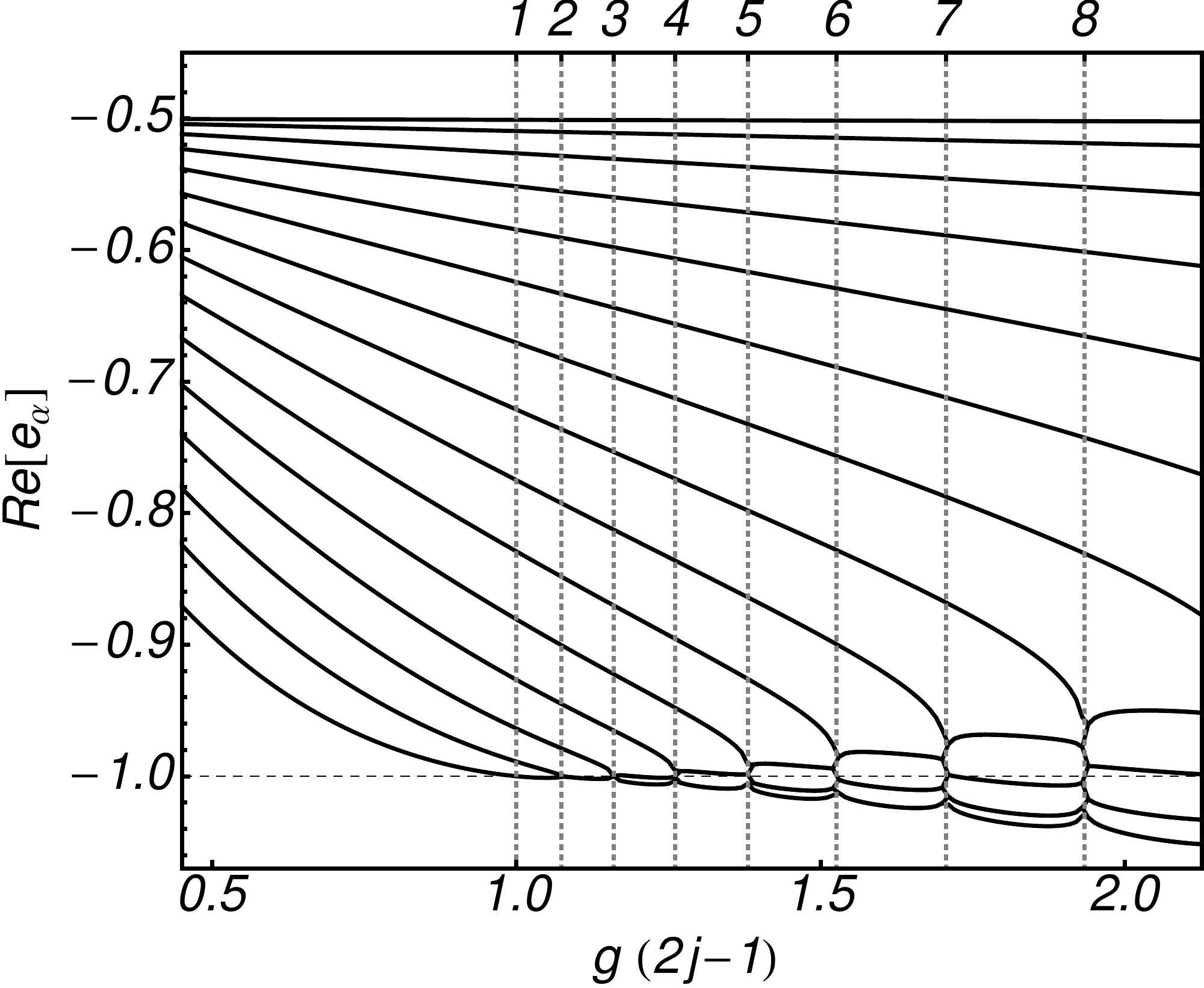}&
\includegraphics[width=0.43\textwidth]{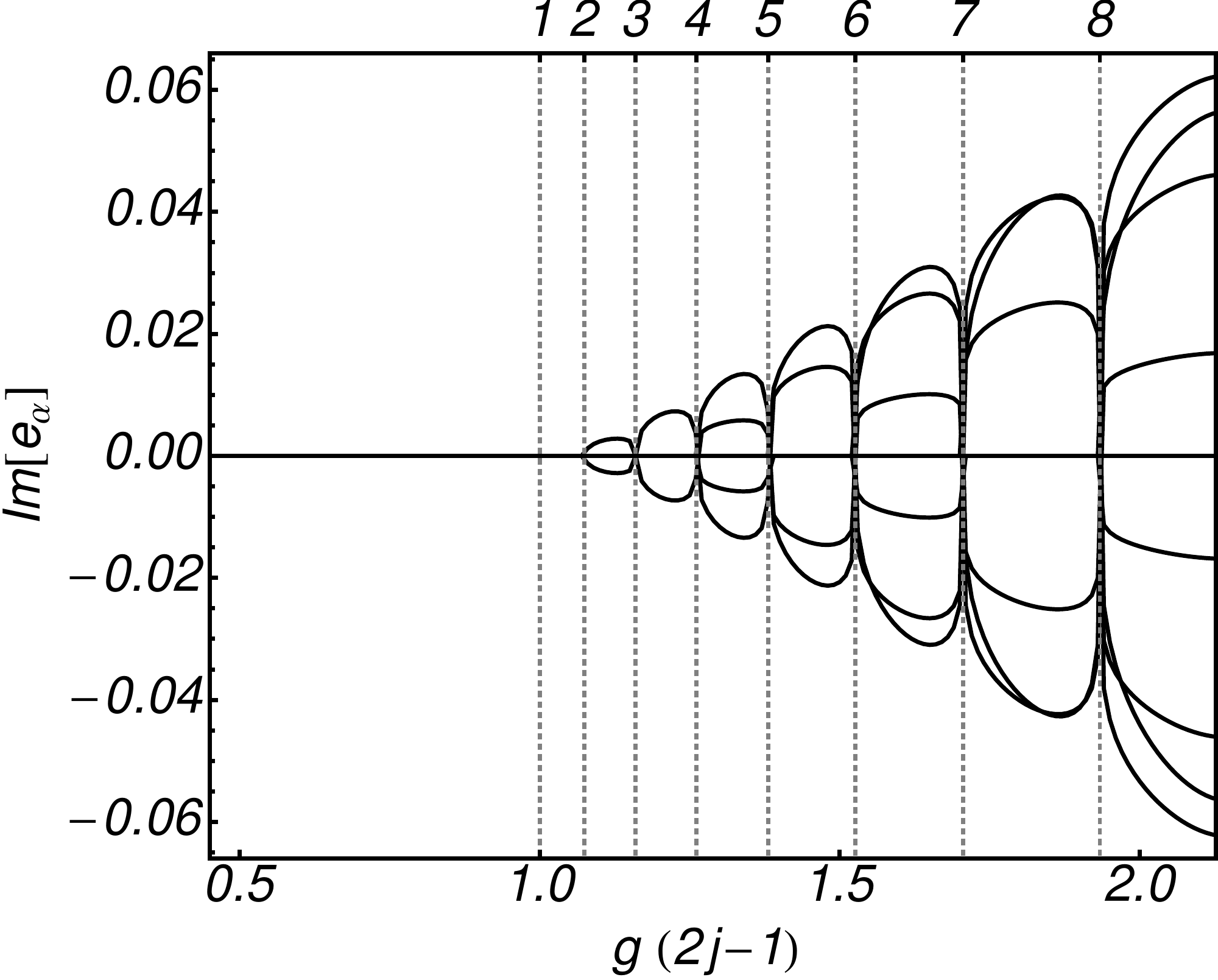}
\end{tabular}
}
\caption{ The collapses of the ground state pairons as a function of  $g(2j-1)$ for the hyperbolic case of figure \ref{fig3} ($t=1/2$ and $j=15$). Left and right  panel show   the real and imaginary part of the pairons respectively. The collapses take place in the position $P_C=-1$ (horizontal dashed line in left panel) of the effective charge $Q_C$ at values of $g$ given in (\ref{cond1}). The dashed vertical lines indicate the  number of pairons $N_C=1,...,8$ involved in the collapse. After the collapse of an even number of pairons a complex conjugated pair is created as can be seen in the right panel of the figure. }
\label{fig4}
\end{figure}

According to this expression, the first collapse occurs for $N_C=1$ (one collapsing pairon)  at $g(2j-1)=g^c_1 (2j-1)=1$, then $N_C=2$ pairons converge to $P_C=-1$ at $g(2j-1)=g^c_2(2j-1)=(2j-1)/(2j-3)$, and so on. After the collapse of an even number of pairons a new complex conjugated pair of pairons is created. In figure \ref{fig4}, the real and imaginary parts of the pairons are shown for positive $g$ values. The successive collapses and creation of complex conjugated pairs can be clearly seen. This behavior was completely unexpected in bosonic RG models,  where pairons  have  always been constrained to the real axis.

A particular situation occurs when all pairons collapse to $P_C=-1$ for $ g=g^c_{N_C=j}=1$. At this particular point the exact ground state eigenstate (\ref{wavfun}) takes the simple form:
$$
|\Psi\rangle_{MR}=\left( \frac{a^\dagger a^\dagger}{t-1}-\frac{b^\dagger b^\dagger}{t+1} \right)^j |0\rangle,
$$
which would be the boson version of the Moore-Read state found for the $p_x + i p_y$ model \cite{Ger1,Links2,Hyp}, derived from the fermionic hyperbolic RG model. Likewise, as in  the fermionic model, the energy of this state is $E=0$.  However, within the LMG model it can be shown by exact diagonalization of very large systems that the collapse of all pairons to $P_C=-1$ is not associated with a ground state phase transition. A subject still under debate for the $p_x + i p_y$ fermionic model \cite{Ger1,Links2,Hyp}.

It is worth  mentioning here  that the condition (\ref{cond1}) of pairons converging to the value $P_C=-1$ applies equally to the excited states,  and  that  similar collapses in the position ($P_D=+1$) of the  effective $Q_D$ charge occurs for excited states in the $g<0$ interval for values given by
\begin{equation}
g=g^c_{N_D}=-\frac{1}{2j+1-2 N_D},
\label{cond2}
\end{equation}
with $N_D=1,...,M$ and $g^c_{N_D=M}<g^c_{N_D=(M-1)}<...<g^c_{N_D=1}<0$.  These issues will be discussed in section \ref{excited}, where it will be  shown that,  even if the collapses are not associated to a ground state phase transition, they are related to the crossings of excited states  of different parities.

\subsection{The  triple point $\gamma_y=\gamma_x=-1$}

\begin{figure}[t*]
\centering{
\begin{tabular}{lr}
\includegraphics[width=0.455\textwidth]{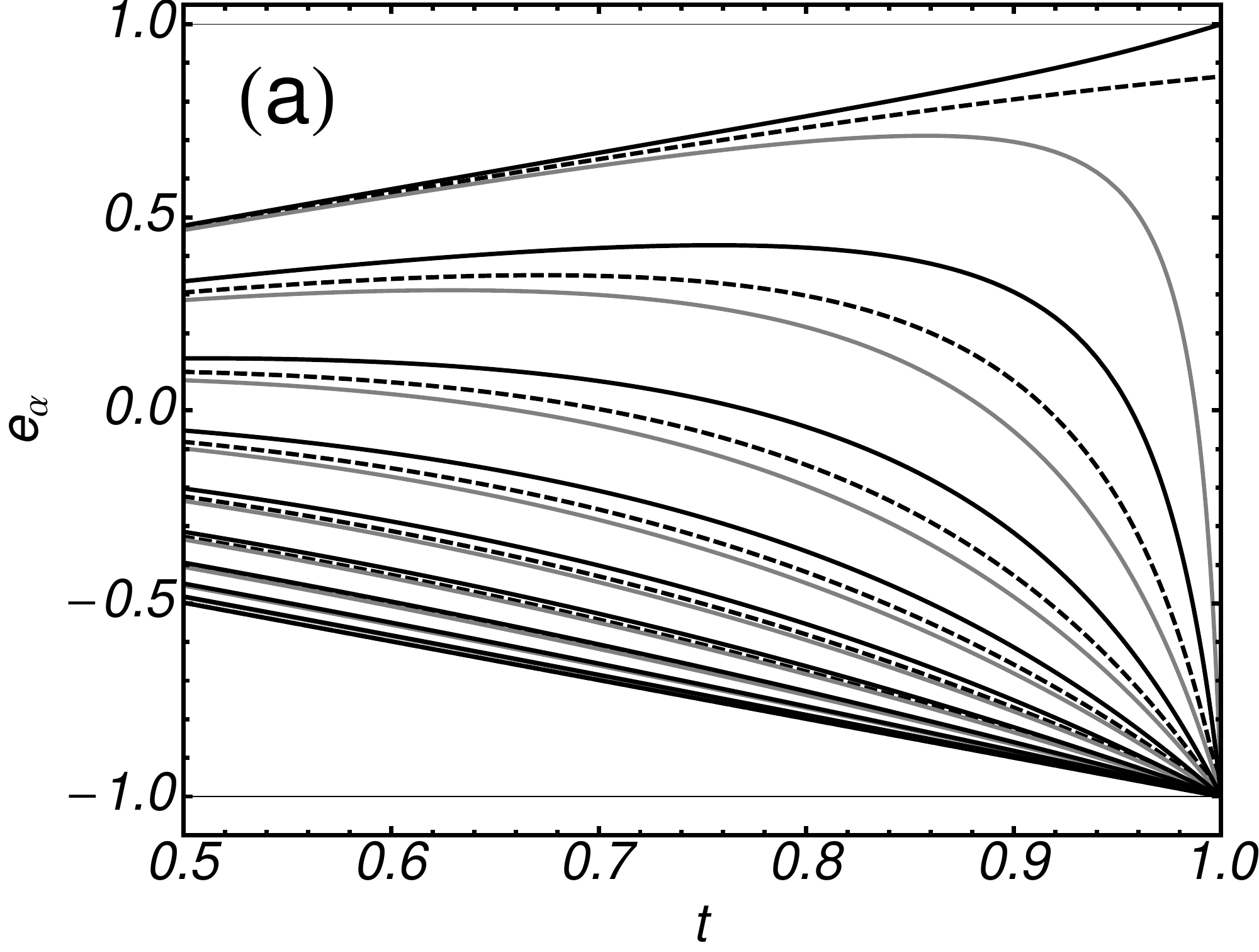}&
\includegraphics[width=0.45\textwidth]{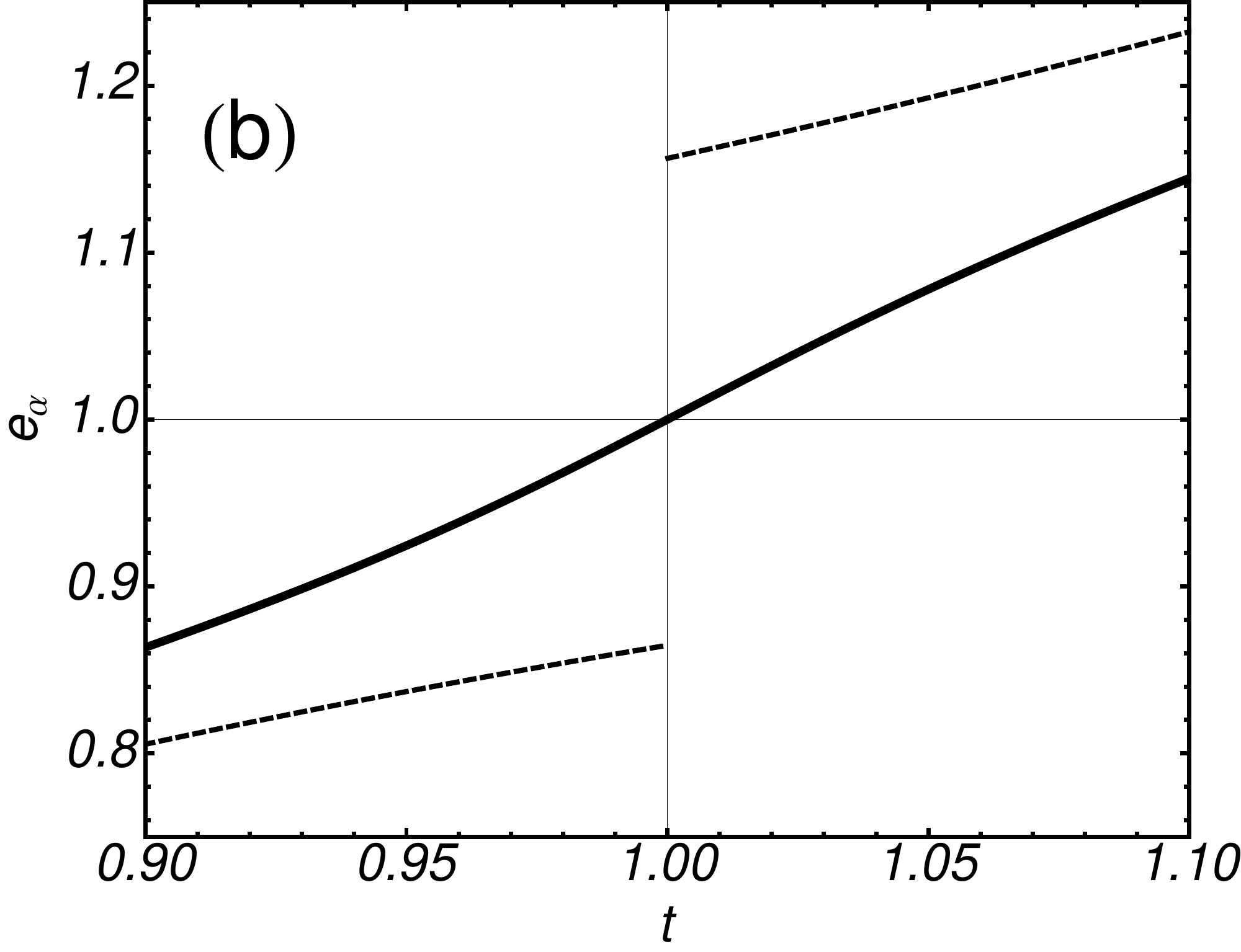}
\end{tabular}}
\caption{Pairons close to the triple point $(\gamma_x,\gamma_y)=(-1,-1)$ for a $j=10$ system as a function of $t$. $t=1$ corresponds to the line $\gamma_x=\gamma_y$ in the phase diagram. The different lines are associated to three values of values of $b$ along the line $\gamma_y=-\gamma_x+b$.  The values $b=-2.0$,  $b_{cr}\approx 2.11$ 
   and $2.32$ are  represented by gray, dashed, and  black   lines respectively, with $b_{cr}=-2 (2j-1)/(2j-2)$.
Panel ({\bf a}) describes the behavior of the ten pairons in the interval $0.5\leq t \leq1.0$. Panel ({\bf b}) is a close up of the tenth pairons around $t=1.0$ showing the discontinuous jump for $b=b_{cr}$.
}

\label{paironsCR}
\end{figure}

The triple point in the phase diagram of the Lipkin model, $(\gamma_x,\gamma_y)=(-1,-1)$, constitutes one of the rare example of a third order phase transition in quantum  many-body systems.  As such, it deserves a thorough study because it could shed light into other third order QPT like the one taking place in the $p_x+i p_y$ model \cite{Hyp}. The third order character of this phase transition reported in \cite{Castanos} is observed when the critical point is crossed, for instance,  along the line $\gamma_y=-\gamma_x-2$. In figure \ref{paironsCR}.a the behavior of the ground state pairons close to the triple point is examined for a system of size $j=10$, moving in the phase diagram along the lines $\gamma_y=-\gamma_x+b$,  for three values of $b$ ($b=-2.0$ grey solid line, $b_{cr}= -2.\overset{\frown}{1}$ dashed line, and  $b=-2.32$ black solid line). We move along these lines  using the parameter  $t$ of the RG model (\ref{gxgy}). The value $t=1$ corresponds to the point $\gamma_y=\gamma_x=b/2$. The critical value of $b$ is $b_{cr}=-2 (2 j-1)/(2j-2)$, which is $b_{cr}<-2$ for finite systems and $b_{cr}\rightarrow -2$ in the thermodynamic limit.   As already discussed, at the points $\gamma_x=\gamma_y$ the LMG Hamiltonian is diagonal in the basis $|jm=(-j + 2k)\rangle\propto (a^ \dagger a^\dagger)^ {(j-k)} (b^\dagger b^\dagger)^{k}|0\rangle$, with eigenvalues given by (\ref{diagE}). The energies $E_{jm=-j}$ and $E_{j m=(-j+2)}$ cross at $\gamma_x=\gamma_y=-(2 j-1)/(2j-2)$. Therefore,  the line $\gamma_y=-\gamma_x+b_{cr}$ traverses a point at which the positive parity states  $|jm=-j\rangle$ and $|j,m=-j+2\rangle$ are degenerated.

Fig. \ref{paironsCR}.a shows the behavior of the  10 pairons in the interval $0.5\leq t \leq1.0$ for the three values of $b$.
In the  three cases the behavior of the nine lowest pairons is similar, all of them converging to $-1$ in the limit $t\rightarrow 1$. However, the last pairon close to $t=1$ distinguishes clearly the three cases studied. For $b>b_{cr}$  the last pairon converges to $-1$ like the other nine pairons, whereas for $b<b_{cr}$ the last pairon converges to $+1$. The critical $b=b_{cr}$ separates both regions. In this case the last pairon converges to $e_-= \frac{\sqrt{j(2j-1)}-1}{\sqrt{j(2j-1)}+1} \approx 0.865$ for $t=1$. The second panel, Fig. \ref{paironsCR}.b, shows more clearly the behavior of the last pairon near $t=1$ for $b=b_{cr}$ and $b<b_{cr}$. Here, the horizontal scale has been extended to $t>1$ using the symmetry transformation (\ref{mirror})  [$(g,t,e_\alpha)\rightarrow (g,1/t,1/e_\alpha)$]. While for $b<b_{cr}$ the last pairon changes continuously across the value of $t=1$, for $b=b_{cr}$ a discontinuity in the last pairon at $t=1$ occurs due to the crossing of positive parity states, jumping from $e_-\approx 0.865$ to $e_+=(1/e_-)\approx 1.156$.
The degeneracy at the point $\gamma_x=\gamma_y=b_{cr}/2$ is associated to the states   $|jm=-j\rangle\propto (a^\dagger a^\dagger)^j|0\rangle$ and  $|j m=(-j+2)\rangle\propto (a^\dagger a^\dagger)^{(j-1)} (b^\dagger b^\dagger) |0\rangle$.
As it can  be inferred from the exact wave function(\ref{wavfun}), the limit $t\rightarrow 1^-$ with $b=b_{cr}$  produces the right eigenstate
 $$|\Psi\rangle=(a^\dagger a^\dagger)^{(j-1)} \left(\frac{a^\dagger a^\dagger}{e_-+1} + \frac{b^\dagger b^\dagger}{e_--1} \right)|0\rangle\propto |\Psi_-\rangle\equiv \frac{ |j m=-j\rangle - |j m=(-j+2)\rangle}{\sqrt{2}}, $$
 which is   a linear combination of the two degenerated states.  The limit  $t\rightarrow 1^+$  (with the last pairon converging to $e_+\approx 1.156$)  produces  the other degenerated state, $|\Psi_+\rangle\equiv(1/\sqrt{2})(|j m=-j\rangle+|j m=(-j+2)\rangle)$,  orthogonal to the previous one. In summary, when the system traverses the line $\gamma_y=\gamma_x$ along the line $\gamma_y=-\gamma_x+ b_{cr}$  the ground state wave function presents a discontinuity, changing from $|\Psi_-\rangle$ to $|\Psi_+\rangle$.

In general,  the first order phase transition along the line $\gamma_y=\gamma_x$ (with $\gamma_x<1$) is due to the crossing of the states  $|jm\rangle$ and $|j (m+2)\rangle$.  As it was illustrated above for the particular case of the states  $|jm=-j\rangle$ and $|j m=(-j+2)\rangle$,  the behavior of the pairons near this line reflects these crossings by a discontinuity in their values. As a result, crossing the line $\gamma_y=\gamma_x$ (with $\gamma_x<1$) implies a jump from a $|\Psi_{-m}\rangle$ ground state to  a $|\Psi_{+m}\rangle$ ground state,  where $|\Psi_{\pm m}\rangle\equiv \frac{1}{\sqrt{2}}(|j m\rangle \pm |j (m+2)\rangle)$. This first order phase transition is signaled by a discontinuous change in the order parameters $\langle S^2_x\rangle$ and $\langle S^2_y\rangle$. It can be shown that in the thermodynamic limit the order parameters for these states  are
\begin{eqnarray}
\frac{\langle \Psi_{\pm m}| S_x^2 |\Psi_{\pm m}\rangle}{j^2}&=& \frac{2\pm 1}{4}\left( 1-\left(\frac{m}{j}\right)^2\right) \nonumber\\
\frac{\langle \Psi_{\pm m}| S_y^2 |\Psi_{\pm m}\rangle}{j^2}&=&\frac{2\mp 1}{4}\left( 1-\left(\frac{m}{j}\right)^2\right).
\end{eqnarray}
Therefore, a jump from $|\Psi_{- m}\rangle$ to $|\Psi_{+m}\rangle$, for $m\not=-j$, produces a discontinuity in the order parameters characterizing a first order phase transition, in complete accord with the analysis of section 1.   For the particular case of figure \ref{paironsCR}  ($m=-j$) both order parameters vanish at the critical point preventing a first order phase transition. The critical value for this continuous phase transition is  $b_{cr}=-2 (2 j-1)/(2j-2)\rightarrow -2$ in the thermodynamic limit, corresponding to the triple point  $\gamma_x=\gamma_y=-1$, in complete agreement  with the thermodynamic results of reference \cite{Castanos}

\section{Excited states in the hyperbolic LMG model}
\label{excited}

We will study in this  section  the excited states  for the hyperbolic regions ($s=-1$) of the LMG model in terms of the  pairon dynamics of the RG model. A similar description for the rational bosonic RG model was performed in \cite{PittDuk}. Pairons for the rational as well as for trigonometric bosonic RG models are always real. The hyperbolic model has the particular feature that the pairons can take complex values. For the sake of clarity, let us assume the specific value $t=1/2$, with generic results for  the cases $t<1$. The effective charges $Q_C$ and $Q_D$ of the Richardson equations (\ref{RichEqsEl}) located  at positions $P_C=-1$ and $P_D=1$,  are outside of the interval $[-t,t]$. The cases with  $t>1$, can be inferred from those with $t<1$ trough the mirror transformation of Eq. (\ref{mirror}).

\begin{figure}[t]
\centering{
\includegraphics
[width=0.5 \textwidth]{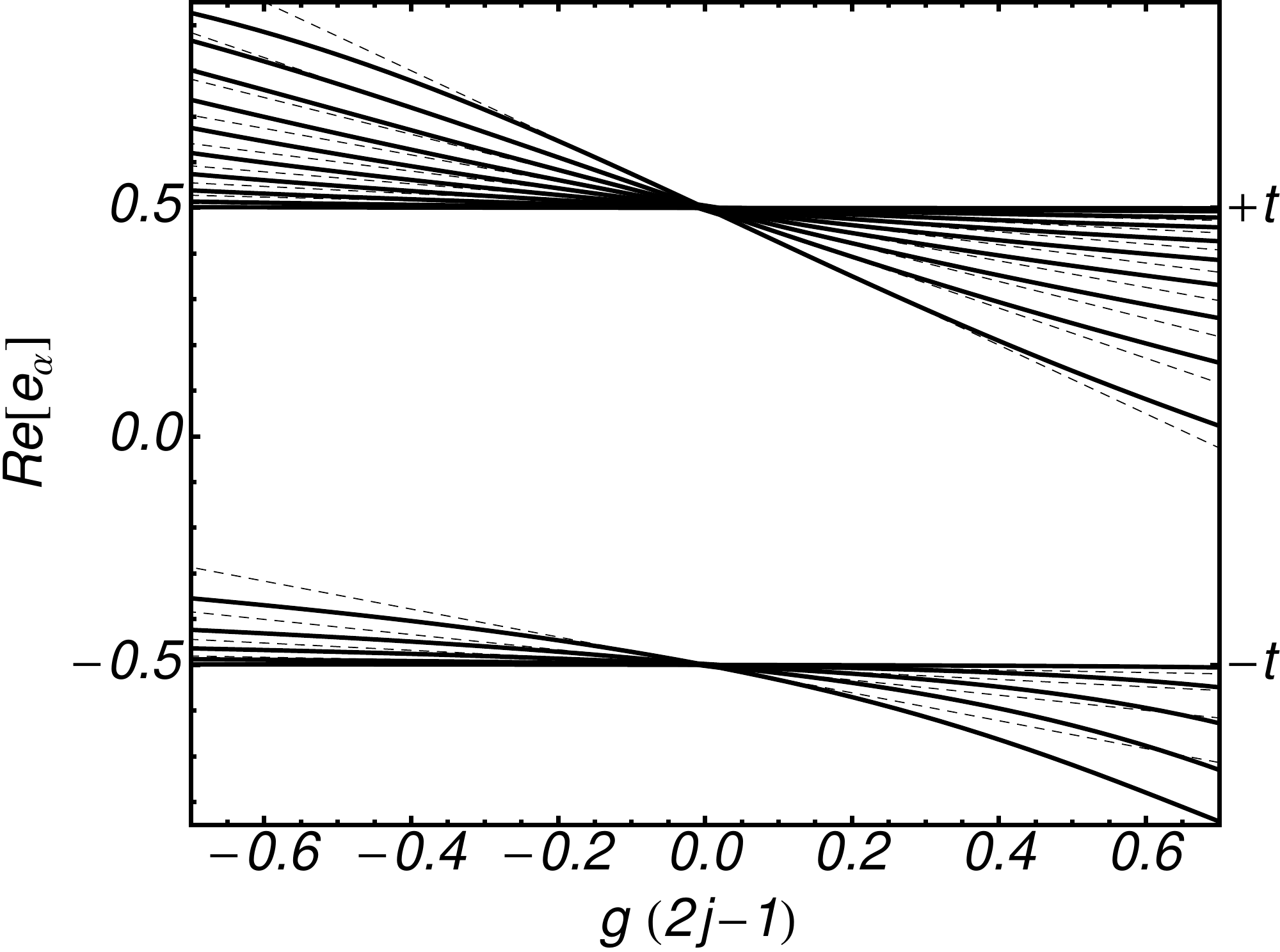}
}

\caption{ Pairons for the  $k=10$-th positive parity excited state of a  $j=15$ system. The perturbative results (\ref{smallg}) are shown by dashed lines and the exact ones  with solid lines. The total number of pairons is $M=j=15$.  The $k=10$-th state is characterized by having $5$ pairons close to $t_a=-t$ and $10$ pairons close to $t_b=+t$ at weak coupling. Note that the lower pairons for negative $g$ are inside  the interval $[-t,t]$, and the upper ones outside. The opposite happens for positive $g$ where the lower pairons are outside the interval, and the upper pairons are inside.}
\label{otra}
\end{figure}

 Let us first analyze the limit $g=0$. In this limit the LMG Hamiltonian reduces to an one-body Hamiltonian $H=\epsilon S_z=\epsilon \frac{b^\dagger b-a^\dagger a}{2}$  with eigenvalues $\epsilon(n_b-n_a)/2$ and eigenstates  $|\Psi\rangle= |n_b=\nu + 2k, n_a=\nu+2(M-k) \rangle$.  Here, the seniorities are the number of  unpaired bosons   $\nu=0,1$, $M=(j-\nu)$ is the total number of  boson pairs, and $k=1,...,M$.  The positive (negative) parity sector corresponds to $n_a$ and $n_b$ even (odd), or in terms of the seniorities it corresponds to $\nu=0$ ($\nu=1$). Independently of the parity, the ground state has $M$ boson pairs occupying the $a$ level. The excited sates are obtained by promoting boson pairs from the $a$ to the $b$ level. In this way the $k$-th excited state for a given parity has $M-k$ boson pairs occupying the $a$ level, and $k$ in the $b$ level. From the wave function (\ref{wavfun}), it can be seen that in the $k$-th excited state $M-k$ pairons  converge to $t_a=-t$ and $k$  pairons to $t_b=+t$.  For  finite but small $g$,  it was  shown in ref.\cite{NPB707} that the  pairons $e_\alpha$ can be approximated by
\begin{equation}
e_\alpha\approx  t_i -g (1+st^2)r_l, \ \ \ \  (s=-1)
\label{smallg}
\end{equation}
where  $t_i=t_a=-t$ or $t_i=t_b=+t$, and  $r_l$ are the positive roots of the Legendre polynomial $L_{N_i}^{\nu-1/2}(x)$, with $N_i$ ($i=a,b$)  the number of $e_\alpha$ pairons converging to $t_i$ for $g=0$, i.e, $N_a=M-k$ and $N_b=k$, for the $k$-th excited state.  Hence, for small $g$, the entire set of states can be classified by the number of pairons distributed close to $t_a$ and close to $t_b$.   For the $k$-th excited state in the limit $g$ negative and small, a group of $M-k$ pairons is close and above $t_a=-t$,  implying  they  are in the interval $[-t,t]$, which will turn be very relevant for their behavior in larger $g$. A second group of $k$ pairons is close and above $t_b=+t$, i.e. outside the interval $[-t,t]$.
For small and positive $g$ the situation is reversed. $M-k$ pairons are  outside the interval $[-t,t]$ and close to $t_a=-t$,  whereas the remaining  $k$ pairons  are inside the interval and close to $t_b=+t$. In Figure \ref{otra} we illustrate this behavior for the positive parity $10$-th excited state of a system with  $M=j=15$ pairons. In the figure,  $k=10$ pairons are close to $t_b=+t$ and the remaining $M-k=5$ sit close to $t_a=-t$.
Even if the perturbative result is not valid for large $g$,  the behaviour of the pairons as a function of $g$ is strongly dependent on  their values at weak coupling. In particular on  whether they are inside or outside the interval $[-t,t]$. The  pairons inside the interval $[-t,t]$ expand  in the real axis but remain constraint to it for any value of $g$. On the contrary, the pairons outside the interval expand on the real axis moving away from the interval. For negative $g$ this expansion takes place above the interval $[-t,t]$ till the pairons  start to collapse in the position $P_D=1$ of the effective charge $Q_D$ for values of $g$ given by (\ref{cond2}).  Similar collapses occur for positive $g$ for values given by (\ref{cond1}), but here the collapses take place at the position $P_C=-1$ of the effective charge $Q_C$.

\begin{figure}[t*]
\centering{
\begin{tabular}{lcr}
\includegraphics[angle=0,width=0.46\textwidth]{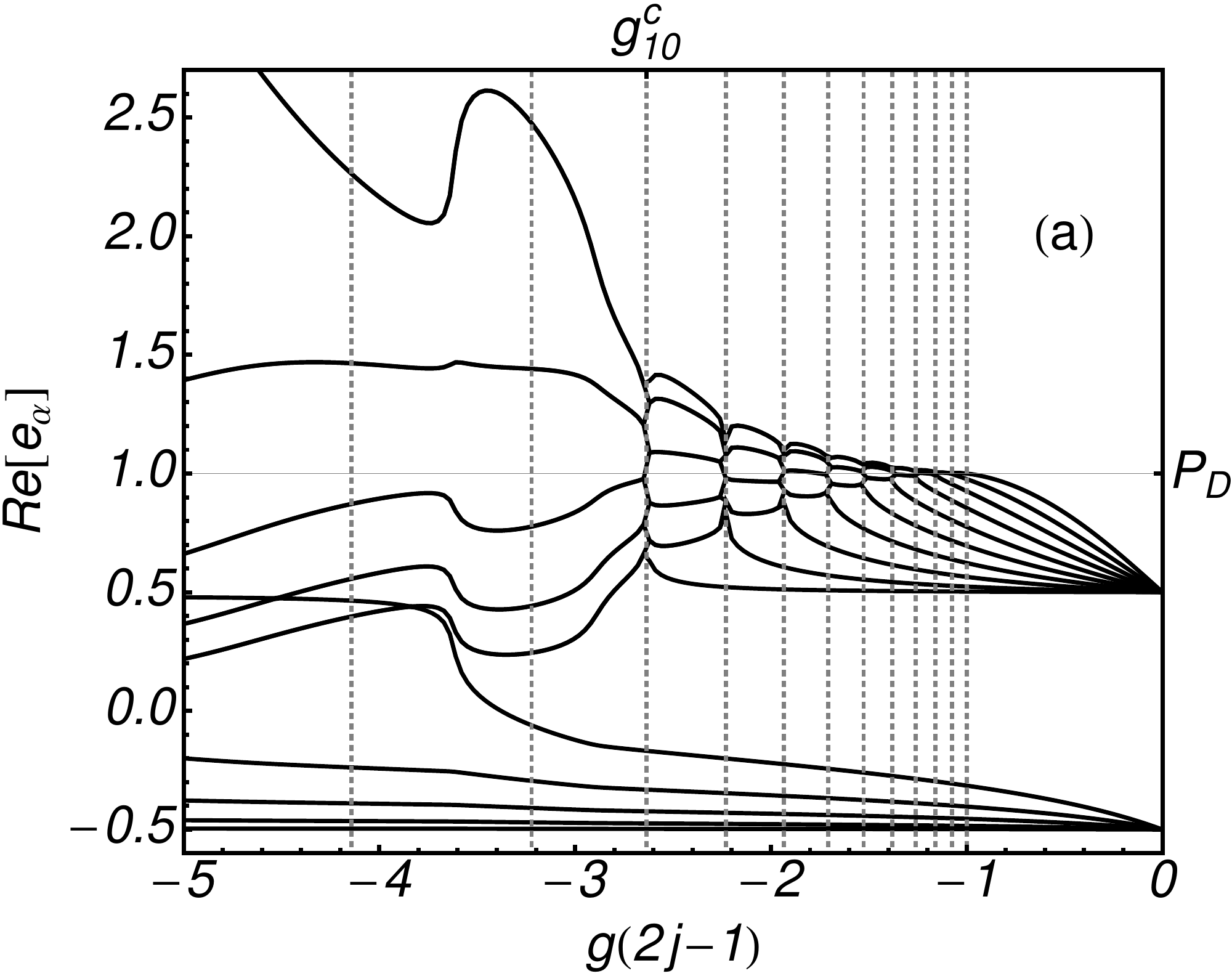} &  &\includegraphics[width=0.44\textwidth]{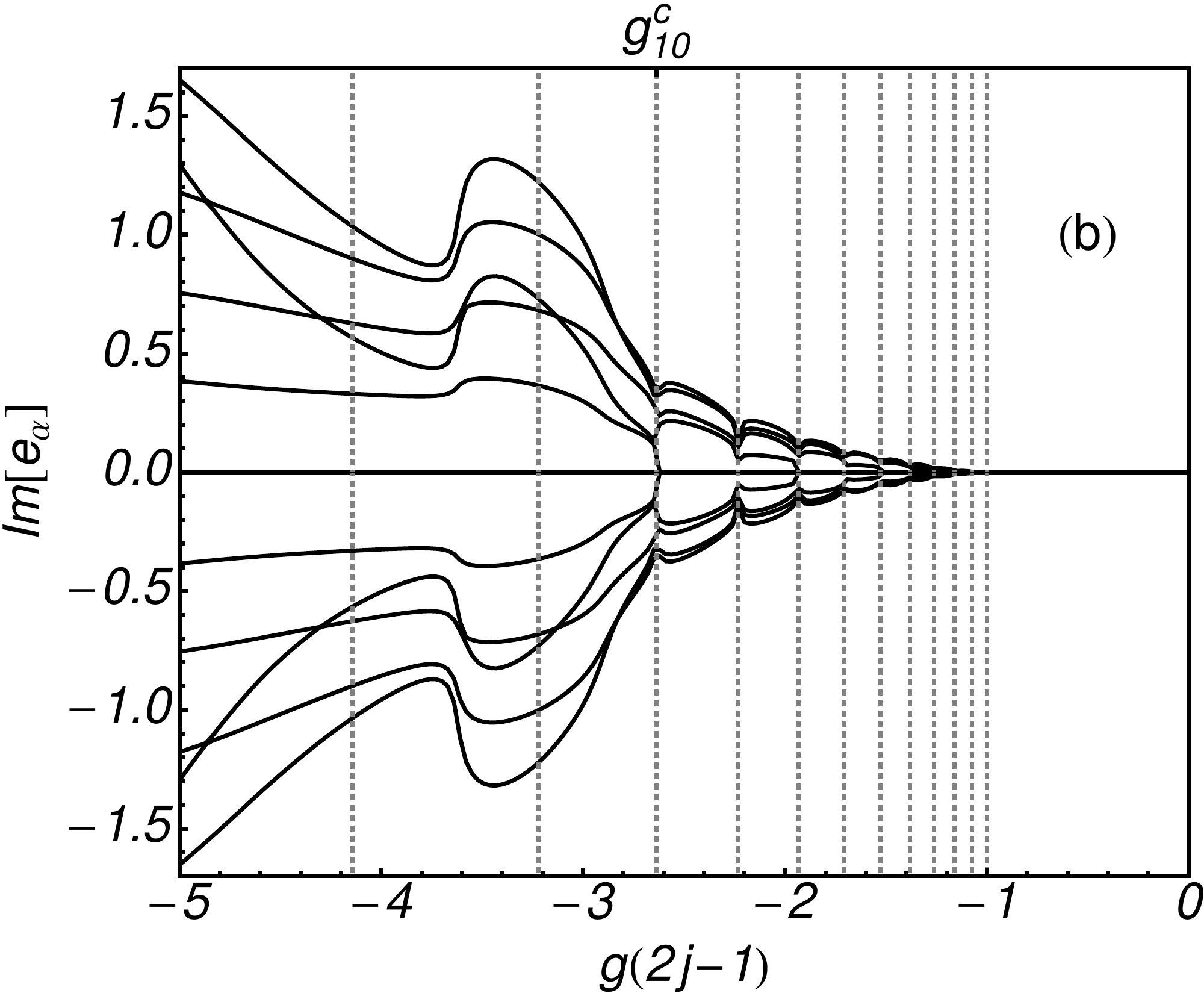}
\end{tabular}
}
\caption{Real ({\bf a}) and Imaginary ({\bf b}) parts of the 15 pairons of the ($k=10$) state of figure  \ref{otra} as function of $g(2j-1)$. Vertical  dashed lines indicate the $g$ values where pairon  collapses occur according to Eq.(\ref{cond2}). The pairons close to $t_a=-t$ in the weak coupling limit remain trapped in the real interval $[-t,t]$, whereas the non trapped pairons expand in the complex plane after collapsing in the position $P_D=1$ [horizontal gray line in panel (a)] of the  effective charge $Q_D$. The number of non trapped pairons for this $10$-th excited state is $10$, therefore collapses of 1 to 10 pairons occur at  $g$ values given, respectively, by  $g=g^c_{N_D}$ with $N_D=1,...,10$. The results corresponding to  $g^c_{N_D=10}$ are indicated in the upper scale of the panels. For $g<g_{N_D=10}$ no more  collapses occur, and the pairon set consist of 5 real pairons in interval $[-t,t]$ and  $5$ complex conjugated pairon pairs. }
\label{repa}
\end{figure}

  For positive values of $g$,  the $k$-th excited state has $M-k$ pairon non trapped in the interval $[-t,t]$.  Therefore, for  the ground state ($k=0$) all the pairons will successively  collapse at $P_C=-1$  for increasing $g$ as it can be seen in figure \ref{fig3}. The first excited state has one pairon trapped in the interval $[-t,t]$, while the other ($M-1$)  non trapped pairons successive collapse at $P_C=-1$  for increasing values of $g$. The same reasoning  extends for the rest of the excited states. For the most excited state,  with  all its pairons  trapped in the interval $[-t,t]$, no collapse occurs.

  The situation is somewhat reversed for negative values of $g$. The $k$-th excited state has $k$  non trapped pairons. Therefore, the higher  the excited state is, the more collapses will occur in the position of effective charge $Q_D$ at  $g$ values  given by Eq. ({\ref{cond2}}).   For the grounds state ($k=0$), since all the pairons are trapped in the interval $[-t,t]$ no collapse occurs, as it can be seen in  figure \ref{fig3}.  For the $k$-th excited state successive collapses of $1$ to $k$ pairons occur at $g$ values given, respectively, by $g^c_{N_D=1}, g^c_{N_D=2},...,g^c_{N_D=k}$.

In Fig. \ref{repa} we illustrate this behavior for the same system and state of figure \ref{otra} ($P=+$, $j=15$ and $10$-th excited state), and for negative values of $g$. For small $g$, five pairons are located close and above  $t_a=-t$ and $k=10$ pairons  are close and above  $t_b=t$. As $g$ is increased the trapped pairons  expand in the interval $[-t,t]$, but remain  constraint to it for  any negative $g$.  Whereas,  the other ten  pairons  expand outside this interval, till they begin to collapse into the position ($P_D=1$) of the effective charge $Q_D$ at $g$ values given by Eq.(\ref{cond2}). These $g$ values are indicated by vertical dashed lines in the  panels of Fig. \ref{repa}. Immediately after the collapse of an even number of pairons two  complex conjugated pairons are created, and they expand in the complex plane until the next collapse. For the case illustrated in the figure, since the number of non trapped pairons is $10$, the last collapse to $P_D=1$ takes place at $g=g^c_{N_D=10}$. From there on  five complex conjugated pairon pairs expand in the complex plane for $g<g^c_{N_D=10}$.

\begin{figure}
\centering{
\begin{tabular}{lr}
\includegraphics[angle=0,width=0.45\textwidth]{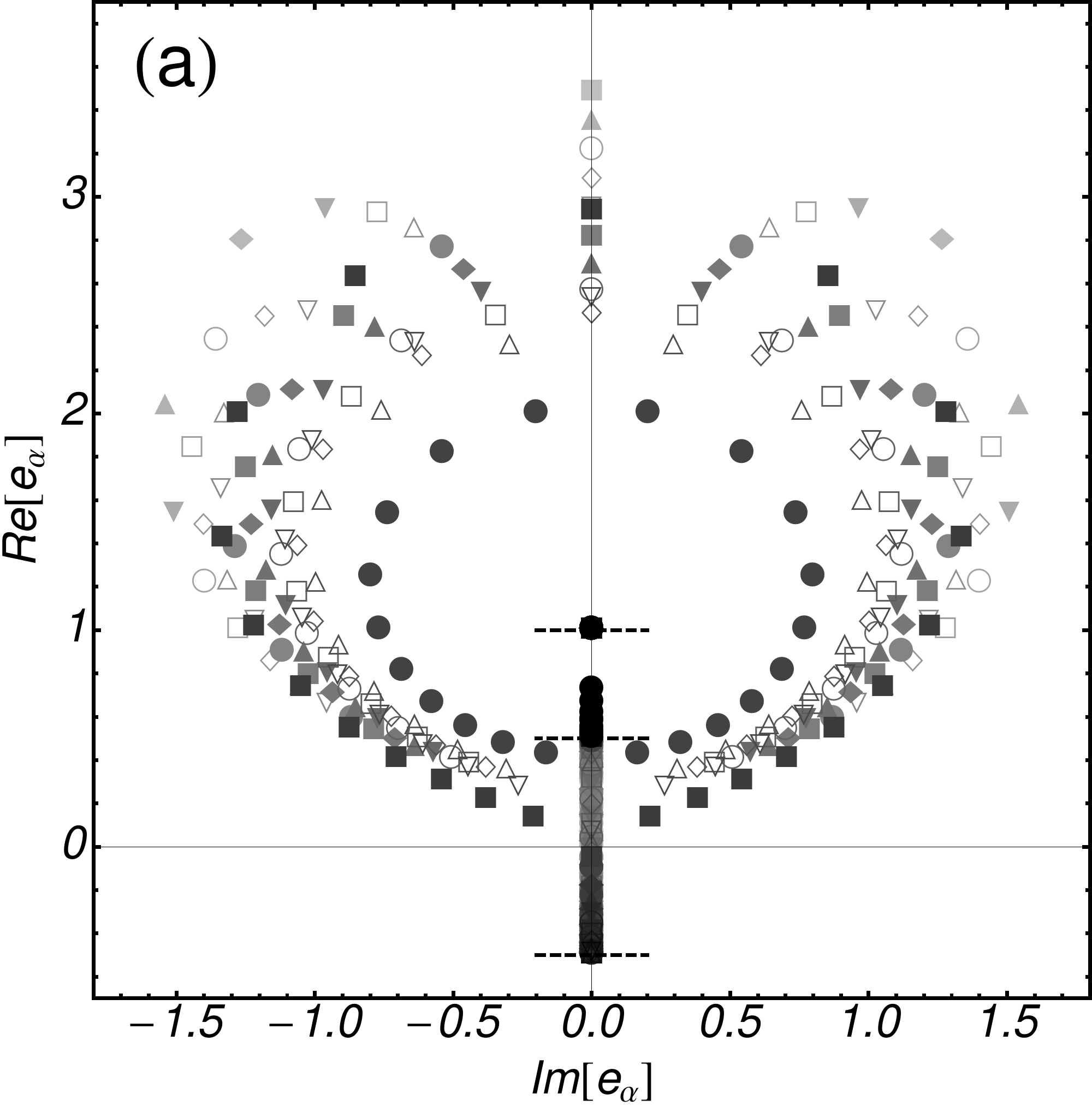}
\includegraphics[angle=0,width=0.45\textwidth]{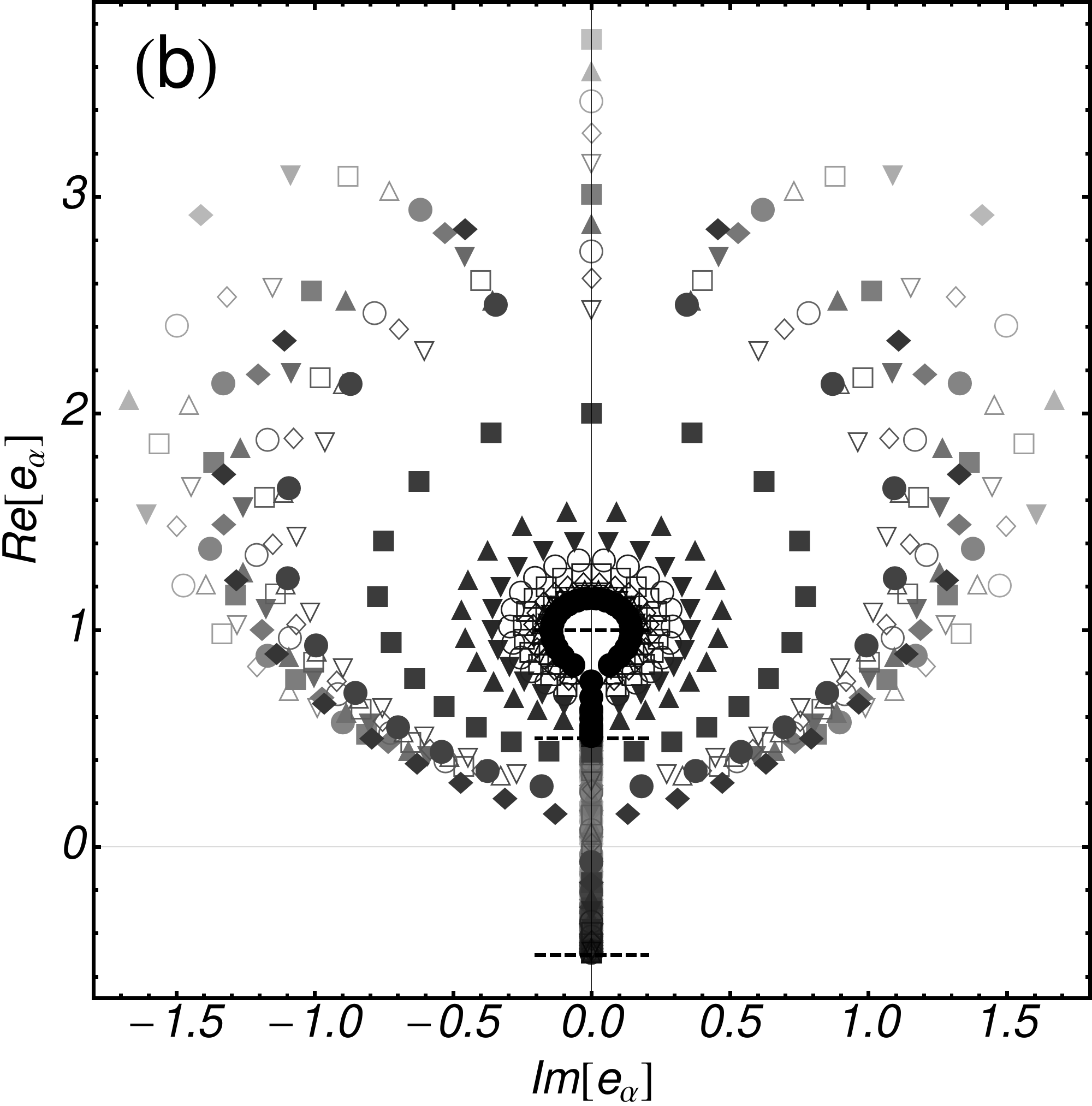}
\end{tabular}
}
\caption{Pairons in the complex plane for the complete set of positive parity  states for a $j=30$ system. Panel ({\bf a}) corresponds to  $g=g_{22}^c$ where collapses of $22$ pairons are  expected according to Eq. (\ref{cond2}). Panel ({\bf b}) corresponds to a value of $g$ where collapses are not expected. Horizontal dotted lines indicate the values $t=-1/2$, $t=1/2$ and $P_D=1$. The  complex pairons of each state accommodate in arcs around  $P_D=1$. Outer arcs correspond to lower energy states. In panel ({\bf a}) the arcs of the $k=22$ to $k=j=30$ excited state collapse to the position $P_D=1$, with the outer arcs corresponding to lower energy excited states.}
\label{purpa}
\end{figure}

A better insight on the pairon dynamics can be gained by plotting the pairon positions in the two dimensional complex plane. Fig. \ref{purpa}  shows the  pairon positions in the complex plane of the complete set of positive parity states for a system of $j=30$ and  two different negative $g$. The fist  one [panel (a)] is $g=g^c_{N_D=22}$ where $N_D=22$ pairons collapse in $P_D=1$.  The second one [panel (b)] is an intermediate value of $g$ between two collapses ($g^c_{N_D=23}<g<g^c_{N_D=22}$).  The complex conjugated pairs of pairons of a given state are distributed in complex arcs around $P_D=1$. In panel (a), the radius of these arcs goes to zero for states with large enough number of non-trapped pairons, i.e. for those states with $k=22$ to $k=j=30$ non trapped pairons, corresponding to the $22$-th to $30$-th excited state. In both cases  the outer arcs correspond to less excited states.

\begin{figure}[t*]
\centering{
\includegraphics[width=.6\textwidth]{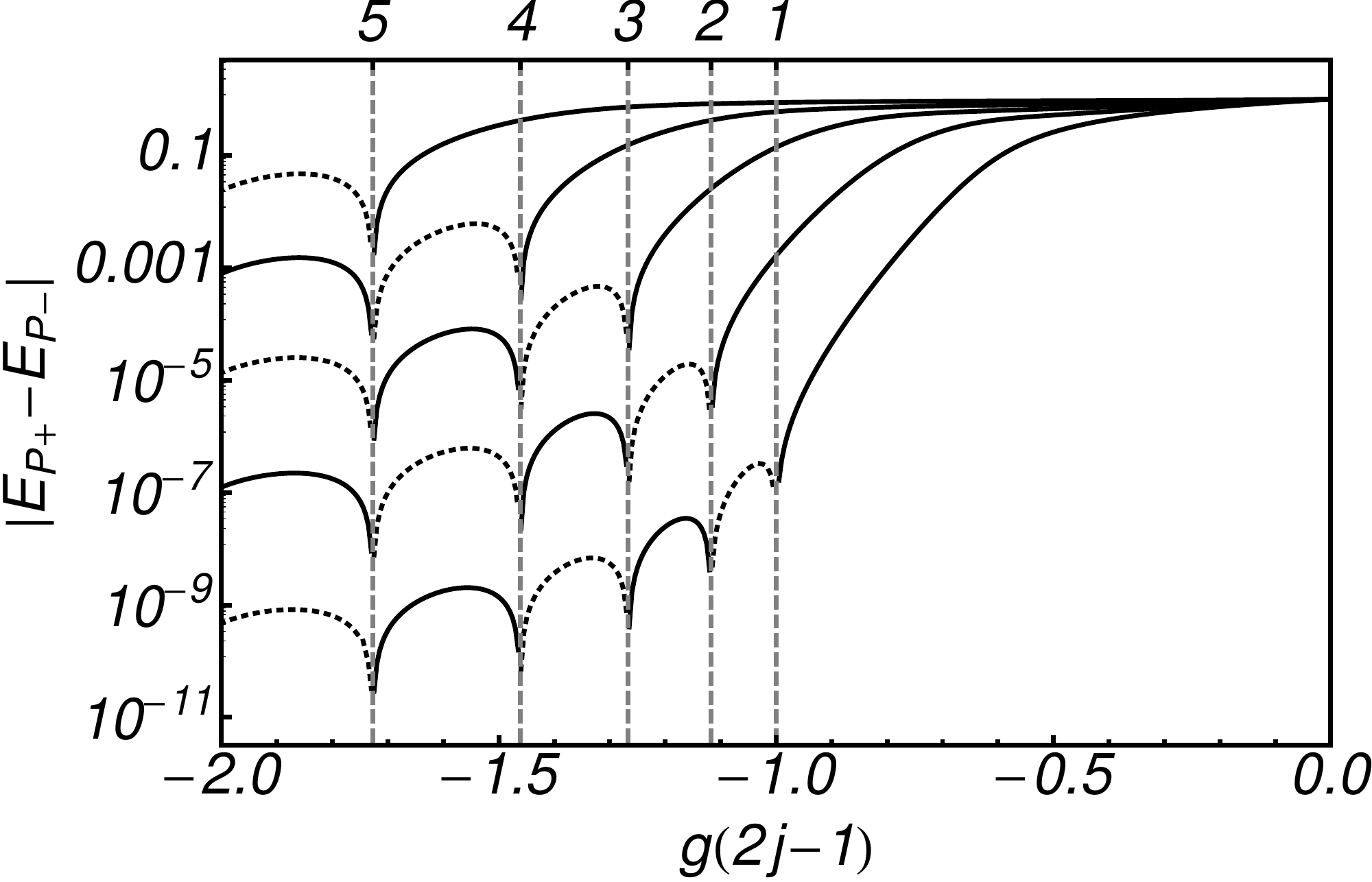}
}
\caption{Absolute value of the difference between the ground to the fourth excited state energies of positive and negative parity states for the hyperbolic LMG model with $j=10$ and $t=1/2$. Continuous segments indicate that the   $P=+$ state is lower in energy respect to the corresponding  $P=-$ state, whereas the dashed segments correspond to the opposite situation $E_{P-}<E_{P+}$. The vertical dashed lines indicate the values $g^c_{N_D}$ (with $N_D=1,...,5$ signaled in the upper scale).}
\label{dife}
\end{figure}

Before closing this section, it is interesting to note that the condition  (\ref{cond2}) of  $N_D$ pairons converging to $P_D=1$ in the hyperbolic region ($s=-1$) of phase diagram,  defines  hyperbolas in the $\gamma_y$-$\gamma_x$ plane given by
\begin{equation}
\gamma_y \gamma_x =\left(\frac{2j-1}{2j+1-2N_{D}}\right)^2.
\label{hyperbolas}
\end{equation}
For $N_D=1$, the resulting hyperbola ($\gamma_y\gamma_x=1$) is the same reported in \cite{Castanos} for the crossing of the ground states of positive and negative parities in the third quadrant. Moreover, in \cite{Castanos2} it is argued that the rest of the hyperbolas ($N_{D}=2,...,M$) define the points where there are crossings between excited states of the two parity sectors. For instance, for $N_D=2$  the ground and first excited state of positive parity cross, respectively, those of the negative parity sector. In general, for arbitrary $N_D$  the corresponding hyperbola defines the points where the first $N_D$ states of positive parity cross, respectively,  the first $N_D$  negative parity states.  This result, already confirmed in \cite{Castanos2,Chen2} for the ground state, is numerically confirmed here  for the ground and excited states  in figure \ref{dife}.  The figure displays the absolute value difference ($|E_{P+}-E_{P-}|$) between $P+$ and $P-$ states  for a system with $j=10$ and $t=1/2$ as a function of  negative $g$ (third quadrant in  the phase diagram  $\gamma_x-\gamma_y$). The differences between the ground, first, second, third and fourth excited states of every parity sector are shown in logarithmic scale in order to make clear the crossings between positive and negative parity states. Every line is divided in continuous and dotted segments indicating if the difference $E_{P+}-E_{P-}$ is  negative or positive respectively. The points where this difference changes sign indicate a crossing between positive and negative parity states. The vertical  dashed lines signal the points where the hyperbolas (\ref{hyperbolas})  are traversed, i.e. when  $g (2j-1)=g^c_{N_D} (2j-1)=-\frac{2j-1}{2j+1-2 N_D}$. As it can be seen in the figure, for $N_D=1$ [where $g(2j-1)=-1$] the ground states of every parity sector cross, whereas for the  leftmost vertical lines ($N_D=2,3,4,5$), in addition to  the ground states, more and more excited states cross. It is worth mentioning that the phase transition in the thermodynamic limit is expected at $g (2j-1)=-t=-1/2$ (see table \ref{tabla}).  We can appreciate in the figure a dramatic change in the difference between the energies of the positive and negative parity ground states around this value. However, the first crossing occurs at $g(2j-1)=-1$.

  A preliminary view to the relation between collapses and crossings of different parity states indicates that the states participating in a given  crossing do not have their pairons collapsing in $P_D=+1$. Contrarily, the pairon collapses of a given state prevent it from having a crossing. As result, the ground state having no pairon collapses, has crossings for all the values $g_{N_D}^c$. By contrast, the most excited state with pairon collapses in every value $g_{N_D}^c$, does not have crossings. For intermediate states, they present crossings in $g=g_{N_D}^c$ only if for this particular  value, their pairons do not present collapses. This condition occurs if $N_D$ is greater than the number of their non trapped pairons ($k$ for the $k$-th excited state). For instance,  the $10$-th $P=+$ excited state of Figure \ref{repa}  cross the $10$-th excited $P=-$ state only for $g_{N_D=11}^c, g_{N_D=12}^c,...,g_{N_D=j}$. Further research in this relation is desirable to establish a deeper connection between both phenomena, crossings and pairing collapses.
 Likewise, the hyperbolic quadrants of the phase diagram have a region where avoiding crossings between states of the same parity take place. This region already identified in \cite{vidal} by studying the density of states in the thermodynamic limit, appears in the hyperbolic case ($s=-1$) when $|g(2j-1)|>1/t$. The relation between avoiding crossings and the pairon behavior in the $SU(1,1)$ RG models  is out of scope of this contribution, and  deserves more work for a complete clarification.

\section{Summary}
\label{summa}

The Schwinger boson representation of the SU(2) algebra allows to connect the LMG model with the two-level SU(1,1) RG  pairing models.  We have exploited this relation to classify the entire parameter space of the LMG model in terms of the three RG families, the rational, the trigonometric and the hyperbolic. This classification 
 sheds new light into the LMG phase diagram and its quantum phase transitions. Moreover, the electrostatic mapping of the trigonometric and hyperbolic models provides new insights into the structure of the different phases.
 We  explored the LMG model from the perspective of the RG models, where the eigenstates are completely determined by the spectral parameters (pairons) of a particular solution  of the non-linear set of Richardson equations. We proposed a numerically robust method to solve the Richardson equations which generalize that of reference \cite{solvPan2}. The method was proven to be suitable for obtaining the pairons of the   complete set of  eigenstates for moderate systems sizes.  

 Using boson coherent states, we have re-derived the  phase diagram  of the LMG model and the characteristics of its different phase transitions. The second order phase transitions were interpreted in terms of the RG solution as a localization-delocalization of the ground state pairons, which takes place when the pairons concentrated around the value $t_a=-t$ at weak coupling, expand in the whole interval $[t_a,t_b]=[-t,t]$.  On the other hand, the first order phase transition was related to the discontinuity of the pairons when the transition line is traversed. This discontinuity was related, in turn, with  the crossings between states of the same parity. We have confirmed that the dynamics of pairons in the rational and trigonometric RG models take place entirely in the real axis. However, it was unexpected to find complex pairon solutions in the hyperbolic regions of the phase diagram.  For the ground sate, complex values of the pairons are obtained after the collapses of an even number of pairons into the position $P_C=-1$ of the effective charge $Q_C$.  It was numerically verified diagonalizing very large systems that no phase transition is associated with this singular pairon behavior, even for the particular case in which all the pairons collapse into $P_C=-1$. For the latter case the ground state wave function has a particularly simple form which is equivalent to the Moore-Read state of the $p_x + i p_y$ model \cite{Ger1, Links2, Hyp}.

A complete classification of the excited states for the hyperbolic regions was given in terms of their pairon positions.  For negative couplings it was found that the $k$-th excited state is characterized by a set of $(M-k)$ pairons trapped in  the real interval $[-t,t]$ while the other  $k$ pairons lie outside this interval. The dynamics of these non trapped pairons show collapses of $N_D$ pairons in the position $P_D=1$ of the effective charge $Q_D$, at  $g$ values given by $g_{N_D}^c=-1/(2j+1-2N_D)$. This singular behavior of the excited state pairons was found to be connected to the crossings of $N_D$ lowest positive parity states with $N_D$ negative parity states, at exactly the same values $g_{N_D}^c$.  As  discussed in Ref. \cite{Angela}, the pairon dynamics can help to identify  significant physical phenomena.  The relation between collapses and crossings is an example of this connection that deserves a deeper study. The collapses of pairons in the positions of the effective charges obtained in the hyperbolic boson RG models discussed here, is a feature also found in the  $p_x+ip_y$ fermion pairing realization of the  hyperbolic $SU(2)$  RG model \cite{Ger1,Links2,Hyp}. In this latter model the collapses were related with another singular phenomenon: a third order phase transition. The insight gained in the study of the LMG model, where the set of pairons of every state in the spectrum is easily accessible, can help to elucidate more intricate mechanisms in other integrable models where the numerical access to the pairon sets is more demanding.

\noindent
\vskip 0.5truecm
{\bf Acknowledgements}\\
J. D. acknowledges support from the Spanish Ministry of Economy and Competitiveness under grant FIS2009-07277.
\noindent
\vskip 0.5truecm


\appendix
\section{Lam\'e differential equation from the  RG equations}
\label{app1}
Here we show that the polynomial $ P(x)=\prod_\alpha^M(x-e_\alpha)$, with $e_\alpha$ being the roots of the Richardson equations, satisfies the  Lam\'e differential equation. The demonstration is based on the simple identity
\begin{equation}
\frac{1}{(x-F)(x-G)}=\frac{1}{F-G}\left(\frac{1}{x-F}-\frac{1}{x-G}\right).
\label{identity}
\end{equation}
From the definition of  $P(x)$ it is straightforward to show that its derivative is
$P'(x)=\sum_{\alpha =1}^{M}\prod_{\beta\not=\alpha} (x-e_\alpha)$.
Therefore, we have the identity
$
\Lambda(x)\equiv P'(x)/P(x)=\sum_{\alpha}\frac{1}{x-e_\alpha}.
$
A derivative of $\Lambda(x)$ yields
$$
\Lambda'(x)=\frac{P''(x)}{P(x)}-\Lambda^2(x)=-\sum_\alpha \frac{1}{(x-e_\alpha)^2}.
$$
From the previous result and the identity (\ref{identity}), we obtain
$$
\frac{P''(x)}{P(x)}=-\sum_\alpha \frac{1}{(x-e_\alpha)^2}+\sum_\alpha\frac{1}{x-e_\alpha}\sum_\beta \frac{1}{x-e_\beta}=\sum_\alpha \frac{2}{x-e_\alpha}\sum_{\beta\not=\alpha}\frac{1}{e_\alpha-e_\beta}.
$$
Now, if $e_\alpha$ are the roots of the Richardson equations, the second sum in the right hand can be substituted according using Eq.(\ref{REF})
$$
\frac{P''(x)}{P(x)}=-\sum_\alpha \frac{2}{x-e_\alpha}\sum_k \frac{\rho_k}{e_\alpha-\eta_k}=\sum_{k}\frac{2\rho_k}{x-\eta_k}\sum_{\alpha} \left(-\frac{1}{x-e_\alpha} + \frac{1}{\eta_k-e_\alpha}\right),
$$
where we have, again, used the identity (\ref{identity}). From the definition of $\Lambda(x)$ the previous equation is equivalent to
$$
\frac{P''(x)}{P(x)}+ \left(\sum_k \frac{2\rho_k}{x-\eta_k} \right)\frac{P'(x)}{P(x)}-\sum_{k}\frac{2\rho_k \Lambda(\eta_k)}{x-\eta_k}=0.
$$
Finally, multiplying the previous expression by $P(x)$ and $A(x)\equiv\prod_k(x-\eta_k)$, we obtain the Lam\'e differential equation (\ref{edo}),

$$
A(x) P''(x)+\left[A(x)\sum_{k=1}^4 \frac{2\rho_k}{x-\eta_k}\right] P'(x)- \Big[\sum_{k} 2 \rho_k \Lambda(\eta_k)\prod_{l\not=k}(x-\eta_l)\Big]P(x)=0.
$$

\section{Condition to have $N_C$ pairons converging in the  position $P_C=-1$ of the effective charge $Q_C$ in the hyperbolic LMG model}
\label{app2}
Let us  assume that $N_c$ pairons converge to $P_C=-1$. Therefore we expand them as $e_\alpha=-1 + \delta z_\alpha $, with $\alpha=1,..., N_C$, and $\delta$ an infinitesimal parameter.  In this limit the Richardson equations corresponding to $\alpha=1,...,N_C$ separate in a term  proportional to $1/\delta$ and terms of order $\mathcal{O}(\delta^0)$. Both terms have to cancel independently. The term proportional to $1/\delta$ reads
$$
\frac{1}{\delta} \left( \frac{Q_C}{z_\alpha}+\sum_{\beta}^{N_c}\frac{1}{z_\alpha-z_\beta}\right)=0\quad {\hbox{      with  }} \alpha=1,...,N_c.
$$
Therefore, in order to have $N_C$ pairons converging to $P_C=-1$ the previous set of equations has to have a solution. Following the lines of \ref{app1}, let us suppose a polynomial whose roots give the solution of the previous set of $N_C$ equations, $P_{N_C}(z)=\prod_{\alpha=1}^{N_C}(z-z_\alpha)$. If the set ${z_\alpha}$  solves the system of  equations, it can be shown that the previous polynomial is a solution of the differential equation
$$
zP_{N_C}''(z)+2 P'_{N_C}(z) Q_C=0.
$$
The general solution is $P_{N_C}(z)=\frac{D z^{-2 Q_C+1}}{1-2 Q_C}+E$, with $D$ and $E$ integration constants. By comparing this solution  with the initial assumption [$P_{N_C}=\prod_{\alpha=1}^{N_C}(z-z_\alpha)$], we obtain the  conditions $D=(1-2Q_C)$ and $-2Q_C+1=N_C$. From latter condition and the definition (\ref{chargesEff}) of $Q_C$  we finally obtain that the necessary condition to have $N_C$ pairons converging into $P_C=-1$ is
$$
g= g_{N_C}^c\equiv\frac{1}{2j+1-2N_C}.
$$
Note that this condition applies to any eigenvector, ground or excited state. Following a similar reasoning it can be shown that the condition to have $N_D$ pairons converging to $P_D=1$ is $-2Q_D+1=N_D$, from where the condition
$$
g= g_{N_D}^a \equiv -\frac{1}{2j+1-2N_D}
$$
follows.
It is important to note that the latter condition is satisfied for negative $g$, whereas the former is for positive $g$.

\end{document}